 \documentclass[preprint,longabstract]{aastex}
\pdfoutput=1

\shorttitle{Search for $\gamma$~Dor and $\delta$~Sct Stars }
\shortauthors{Bradley et al.}

\begin{document}


\title{Results of a Search for $\gamma$~Dor and $\delta$~Sct Stars 
with the {\it Kepler} spacecraft}


\author{P.A. Bradley}
\affil{XCP-6, MS F-699 Los Alamos National Laboratory,
    Los Alamos, NM 87545}
    \email{pbradley@lanl.gov}

\author{J.A. Guzik}
\affil{XTD-NTA, MS T-086 Los Alamos National Laboratory,
    Los Alamos, NM 87545}
    
\author{L.F. Miles}
\affil{XCP-6, MS F-699 Los Alamos National Laboratory,
    Los Alamos, NM 87545}

\author{K. Uytterhoeven}
\affil{Instituto de Astrofisica de Canarias, 38200 La Laguna, Tenerife, Spain and 
Departamento de Astrofisica, Universidad de La Laguna, 38200 La Laguna, 
Tenerife, Spain}

\author{J. Jackiewicz}
\affil{New Mexico State University, Las Cruces, NM 88003}

\and

\author{K. Kinemuchi}
\affil{Apache Point Observatory, Sunspot, NM 88349}




\begin{abstract}
The light curves of 2768 stars with effective temperatures and surface gravities placing them near the 
gamma Doradus/delta Scuti instability region were observed as part 
of the Kepler Guest Observer program from Cycles 1 through 5.
The light curves were analyzed in a uniform manner to search for gamma Doradus, delta Scuti, and hybrid star pulsations. 
The gamma Doradus, delta Scuti, and hybrid star pulsations extend asteroseismology to stars slightly more
massive ($1.4$ to $2.5$~M$\odot$) than our Sun. 
We find 207 gamma Doradus, 84 delta Scuti, and 32 hybrid candidate stars. 
Many of these  stars are cooler than the red edge of the gamma Doradus instability strip 
as determined from ground-based observations made before {\it Kepler}. 
A few of our gamma Doradus candidate stars lie on the hot side of the ground-based 
gamma Doradus instability strip. 
The hybrid candidate stars cover the entire region between 6200 K and 
the blue edge of the ground-based delta Scuti instability strip. 
None of our candidate stars are hotter than the hot edge of the ground-based delta Scuti instability strip. 
Our discoveries, coupled with the work of others, shows that {\it Kepler} has discovered over 2000 gamma Doradus, 
delta Scuti, and hybrid star candidates in the 116 square degree {\it Kepler} field of view. 
We found relatively few variable stars fainter than magnitude 15, which may be because they 
are far enough away to lie between spiral arms in our Galaxy, where there would be fewer stars. 

\end{abstract}



\section{Introduction}

The {\it Kepler} spacecraft \citep{bor10,koch10} was launched on 2009 March 6
with the primary goal of searching for extrasolar planet transits. However, the micromagnitude
 precision of the {\it Kepler} light curves make {\it Kepler} observations ideal for finding new pulsating variable stars. 
 Two classes of variable stars with spectral types A and F are of particular interest to us. 
 These are the gamma Doradus ($\gamma$~Dor) and delta Scuti ($\delta$~Sct) stars. 
 They are slightly more massive ($1.4$ to $2.5$ solar masses) and slightly hotter 
 (effective temperature $T_{\rm eff}=$ 6500 to 8500~K) than our Sun. 
 The known $\gamma$~Dor stars pulsate with periods from 0.3 to 3 days and the 
 gravity modes (g modes) are driven by the Òconvective blockingÓ mechanism \citep{guzik00,dupret04,gri05}. 
 They are generally cooler than $\delta$~Sct stars, with a $T_{\rm eff}$ between 6500 and 7500 K. 
 $\delta$~Sct star pulsations are low-order pressure modes (p modes) and mixed character modes (displaying 
 p~mode and  g~mode properties) driven by the kappa, gamma mechanism acting in the He II ionization region. 
 Their periods range from 30 minutes to 
 5 hours, and the $T_{\rm eff}$ values are typically between 7000 and 8000 K. 
 The overlap of $T_{\rm eff}$ and surface gravity ($\log$~g) ranges for these two types of variables 
 suggests the possibility that some 
 stars might show both $\gamma$~Dor and $\delta$~Sct (so-called ÒhybridÓ) pulsation behavior. 
 This hybrid behavior is especially exciting for asteroseismology, since the g~modes of 
 $\gamma$~Dor stars sample deeper regions of the star than the p~modes and mixed modes (with p and g~mode
 properties) of  $\delta$~Sct stars. 
 Ground-based observations discovered only four \citep{henfec05,uytt08,han09} ÒhybridÓ stars. 
 The gamma Doradus ($\gamma$~Dor), delta Scuti ($\delta$~Sct) and ``hybrid'' stars span the transition
 between lower mass stars with radiative cores, relatively deep convective envelopes, 
 and solar-like oscillations (like the Sun) and higher mass stars 
 that have convective cores and radiative envelopes (such as $\beta$~Cephei and slowly pulsating B stars). 
 Current pulsation theory (Guzik et al. 2000; Dupret et al. 2004) shows that there is a complicated energy flow
 regulation by the convection zone and helium partial ionization zones that determines what type of pulsation
 mode a given $\gamma$~Dor, $\delta$~Sct, or hybrid star can pulsate in. 
 Grigahc\'ene et al. (2010) and Uytterhoeven et al. (2011) found from {\it Kepler} data many more hybrid star
 candidates and that they occupy a broader region of the $T_{\rm eff}$, $\log$~g space than current stellar pulsation 
 theory predicts. 
 This means that our understanding is incomplete on the structure of the outer layers ($\delta$~Sct star
 pulsations probe this region) and the deeper layers between the convective core and 
 outer convection zone ($\gamma$~Dor star pulsations probe this region).
 Alternatively, our understanding of the driving mechanisms of these stars is incomplete. 
 
 However, $\gamma$~Dor-type pulsations are hard to observe from the ground 
 due to the relatively low amplitude and approximately one day period of the pulsations. 
 Space-borne telescopes do not suffer these limitations and the MOST satellite discovered two 
 bright hybrid candidates \citep{king06,rowe06}. 
 The larger telescope of the 
 COROT satellite revealed more candidate hybrid stars \citep{har10}. 
 As mentioned above, the initial data sample from the {\it Kepler} spacecraft 
 \citep{gri10} revealed that many of the $\delta$~Sct and $\gamma$~Dor 
 stars found were in fact, hybrid star candidates.
 However, some of the low frequency modes may possibly be Nyquist reflections of high frequency modes \citep{murph13}. 
 Uytterhoeven et al. (2011) followed up by studying a large ($>750$ star) sample of 
 $\gamma$~Dor/$\delta$~Sct candidates that had been labelled as $\gamma$~Dor or 
 $\delta$~Sct candidates by the {\it Kepler} Asteroseismic Science Operations Center (KASOC).
 Of the 471 stars that exhibited 
 $\delta$~Sct or $\gamma$~Dor pulsations, 36{\%} (171 stars) were hybrid star candidates. 
The {\it Kepler} spacecraft also discovered many non-hybrid $\gamma$~Dor and 
 $\delta$~Sct star candidates. 
 Balona et al. (2011)  examined over 10,000 {\it Kepler} stars for $\gamma$~Dor pulsations. 
 They found 137 stars with asymmetric light curves (characterized by strong beating 
 and one or two dominant peaks in the Fourier Transform (FT)). 
 Another 1035 stars were approximately symmetric with respect to their maxima and minima, 
 and finally there were 108 stars with many peaks of relatively low amplitude. 
 Some of the stars that have periods consistent with $\gamma$~Dor variability have 
 temperatures that lie outside the known instability strip. 
 Many of these stars possibly have rotating starspots instead, but spectroscopic data would be needed to 
 confirm this. 
 Tkachenko et al. (2013) screened the publicly available quarter 0 through quarter 8 data, along with
 their Guest Observer data to search for $\gamma$~Dor variability and supplemented this with spectroscopic
 data to confirm their identification as $\gamma$~Dor pulsators and locate them more accurately in an H--R diagram. 
 Balona and Dziembowski (2011) examined over 12,000 {\it Kepler} stars for $\delta$~Sct variability 
 and found 1568 $\delta$~Sct candidate stars. 
 They found that the maximum amplitude distribution increases 
 towards smaller amplitudes. 
 That is, many more $\delta$~Sct stars have their largest modes below 100 ppm rather than above 1000 ppm. 
 They also found that no more than 50{\%} of the stars within the $\delta$~Sct temperature 
 range were variable, which implies that temperature alone is not the deciding factor for 
 $\delta$~Sct pulsations, and that other factors, such as evolutionary state or surface gravity 
 may also play a role in $\delta$~Sct variability.
 
 In summary, the previous satellite surveys show several things. 
 First, $\gamma$~Dor candidates are relatively numerous. 
 The relatively small number 
 of {\it bona fide} $\gamma$~Dor stars discovered with ground based observations was the result of 
 selection effects due to their small amplitudes and their periods being near 1 day.
 Periods near 1 day are very difficult  to discern from the ground due to aliasing and transparency variations. 
 Second, hybrid stars are common. 
 Finally, the large number of modes seen in some stars \citep{por09,gar09,chap11} shows that higher {$\ell$} modes are 
 present at low amplitudes. 
 We collected Guest Observer data from quarters 1 through 17 of a large sample of previously unobserved stars that were
 mostly fainter than magnitude 14. 
 Our goal was to discover candidate $\gamma$~Dor, $\delta$~Sct, and hybrid stars and
 determine their $T_{\rm eff}$ and $\log$~g distribution.
 We examined stars that were within and cooler than the ground based instability strip.
 In particular, we wanted to answer several questions. 
 First, do the new candidate stars lie within the previously established ground-based instability strip boundaries?
 Second, what is the magnitude limit of {\it Kepler}'s ability to detect variable star candidates?
 Finally, what is the relative frequency of $\gamma$~Dor, $\delta$~Sct, and hybrid stars at faint magnitudes
 and how does it compare to other observations involving brighter stars and/or previously known candidates?

 The rest of the paper is organized as follows. 
 We describe our {\it Kepler} dataset in more detail in Section 2. 
 Section 3 provides an overview of our frequency analysis. 
 We discuss the frequency spectra of our target stars in Section 4 and we summarize 
 our findings in Section 5.

\section{Data}

The {\it Kepler} spacecraft monitors the brightness variations of about 150,000 stars in a 
$10^5$ square degree field between Cygnus and Lyra. 
Brown et al. (2011) obtained multi-band photometry (griz, DD051, and JHK from the
Two Micron All Sky Survey) for more than 4 million stars in the {\it Kepler} field. 
With these colors, $T_{\rm eff}$, $\log$~g, visual absorption ($A_v$), 
metallicity, and radius were derived. 
About 150,000 of these stars were selected for continuous monitoring by the {\it Kepler} spacecraft. 
The observations were obtained with a single filter whose effective wavelength 
is close to that of the Johnson R filter \citep{koch10}. 
The magnitudes from this bandpass are referred to as {\it Kepler} magnitudes ($K_p$) in the 
{\it Kepler} Input Catalog (KIC). 
The light curves were obtained in short cadence (1 minute \citep{gil10}) or 
long cadence (30 minutes \citep{jenk10}). 
Each quarter, the stars monitored can change as the {\it Kepler} planet search team refines their 
target selection and new objects are added by participants in the {\it Kepler} Guest Observer program. 
All {\it Kepler} stars are given a unique identification (KIC number), which we use in this paper.

In our study, we select stars whose KIC parameters imply a location in or near the $\delta$~Sct 
and $\gamma$~Dor instability strips, that is, late A to mid F spectral types. 
Our Guest Observer (GO) sample was limited on the bright side by all the stars brighter than 
14.0 being reserved for observations by others (especially the planet transit mission). 
The faintness limit and contamination factor were set to keep the number of stars requested at a 
manageable level (a few thousand). 
The effective temperature range is $8200 > T_{\rm eff} > 6200$~K, and the $\log$~g range is $3.8$ to $4.5$. 
We chose a wider range of $T_{\rm eff}$ and $\log$~g than has been seen in the ground-based instability 
strip to account for the $\pm 250$~K effective temperature and $\pm 0.25$~dex $\log$~g measurement 
uncertainties \citep{mol11,uytt11}. 
We also wanted to ensure that we considered stars outside the ground-based instability strip, 
especially on the cool side in order to confirm that the red edge of the instability strip is not the result
of ground-based sensitivity limits. 
Our selection criteria, coupled with the brighter stars ($<$ magnitude 14) near 8000~K being observed by others, 
resulted in relatively few stars in our sample near the $\delta$~Sct blue edge, 
as shown in Figure~1.
As a result, we expect a significant fraction of our stars will be nonvariable, because they lie 
outside the ground-based instability strip, which we initially assume to apply to the space-based data. 

In the end, we observed 2768 stars with {\it Kepler} as part of the GO program. 
Pinsonneault et al. (2012) showed that the KIC temperatures are actually about 200~K higher 
than believed, and we applied this shift to select the star sample in Cycle 4 (Quarter~13-Quarter~17). 
Our sample stars are shown in Figure~1, along with the ground-based instability strips. 
 The stars of Cycle 2 and 3 were selected based on limiting magnitude (mag $< 15.5$) 
 and contamination factor cutoff ($<10^{-3}$ for Cycle 2 and $< 10^{-2}$ for Cycle 3) . 
 The stars from Cycle 4 had a higher contamination factor limit ($<0.05$) and in addition showed 
 variability in sequences of full-frame images taken in Cycle 0 \citep{kin11}. 
 With few exceptions, we chose no stars brighter than magnitude $K_p < 14.0$ or fainter than 
 $16.0$ (the few exceptions will be discussed separately). 
 Almost all of the stars lie between $K_p = 14.0$ and $15.5$ (see Figure~2 for a magnitude distribution).
 The $T_{\rm eff}$ and $\log$~g values were mostly derived from photometry for all of our program stars, 
 and hence the observational error bars are relatively large ({$\pm 250$}~K and {$\pm 0.25$}~dex for 
 $T_{\rm eff}$ and $\log$~g, respectively) \citep{mol11,uytt11}.  

\begin{figure}
\epsscale{.80}
\plotone{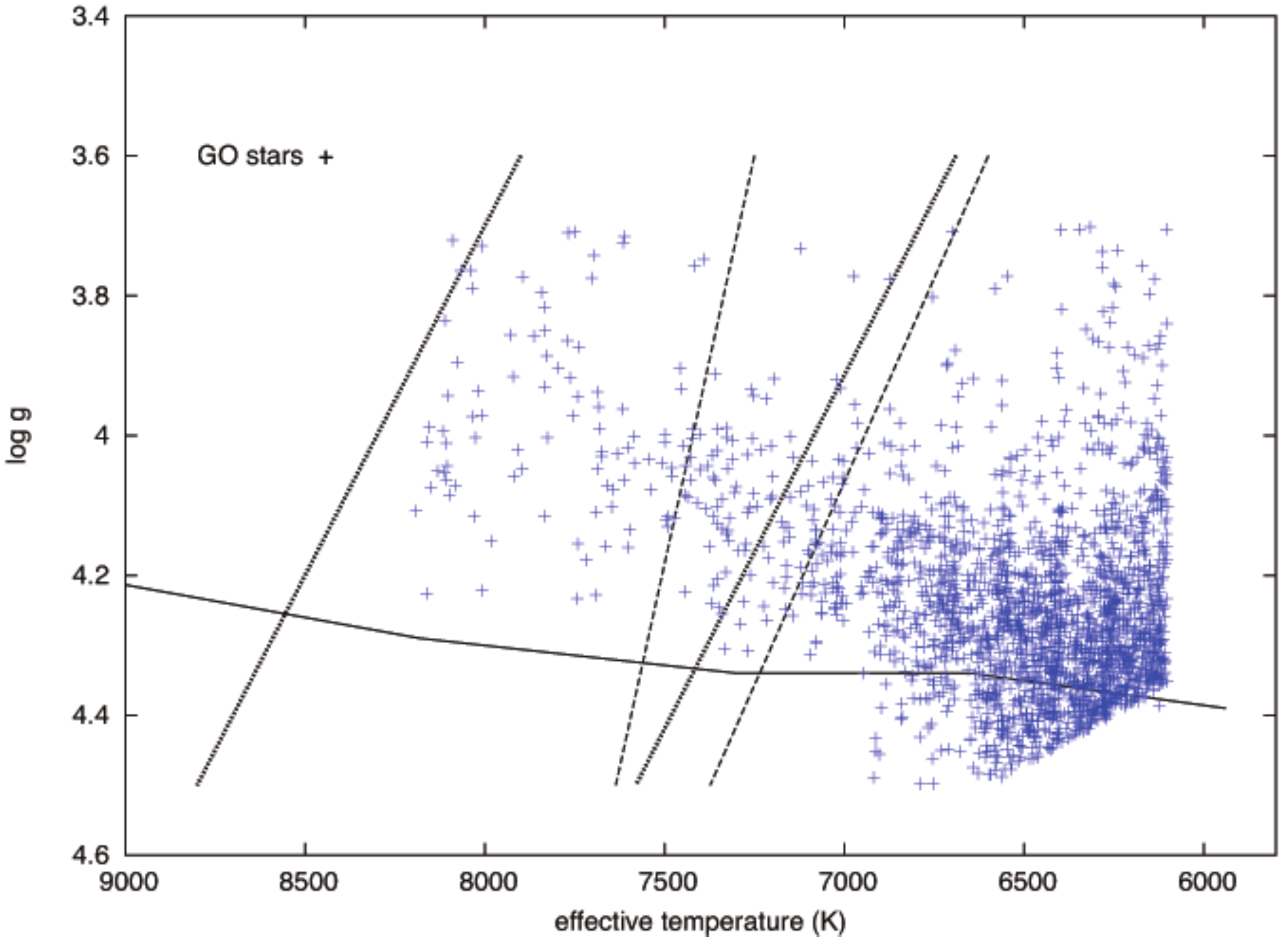}
\caption{Location of our sample stars shifted by +200 K (labeled ÒGOÓ) in the $T_{\rm eff }$, $\log$~g diagram. 
The ground-based $\delta$~Sct (thick dotted lines; Rodriguez and Breger 2001) and $\gamma$~Dor 
(thin dashed lines; Handler and Shobbrook 2002) instability strips are indicated, along with the zero-age 
main sequence (solid line; Cox 2000). Note that relatively few stars lie within the ground-based 
$\delta$~Sct and $\gamma$~Dor instability strips.\label{fig1}}
\end{figure}

 \begin{figure}
\epsscale{.80}
\plotone{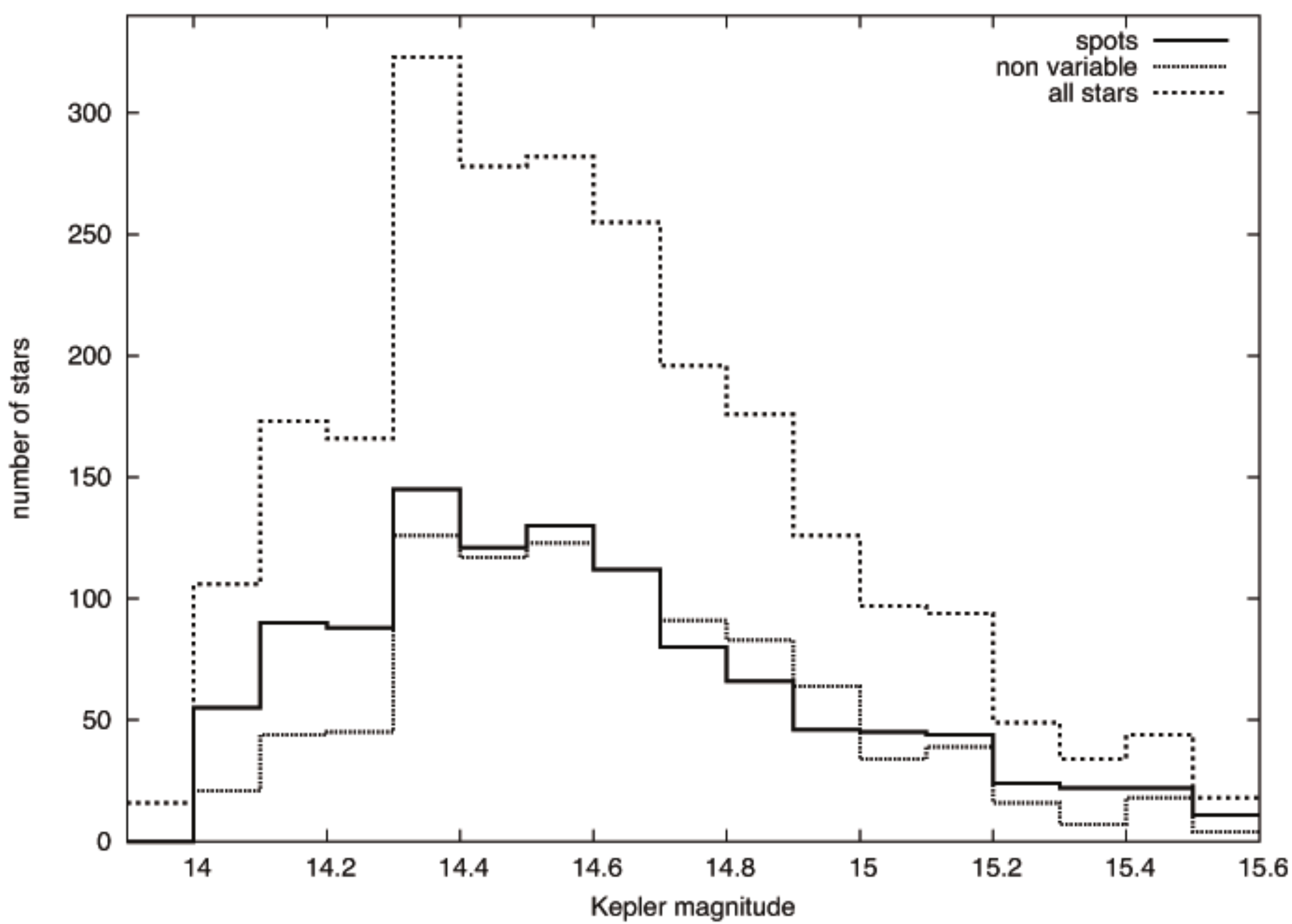}
\caption{Histogram of our sample stars binned by {\it Kepler} magnitude ($K_p$). 
We include the entire sample (dashed line), the non-variable stars (dotted line), 
and rotating stars with spots (solid line). Stars are considered to be non-variable if there 
are no peaks visible in the FT greater than 30 $\mu$mag. 
The number of stars reaches a peak near 
magnitude 14.4 and drops steadily towards fainter magnitudes.\label{fig2}}
\end{figure}


 In addition to the Cycle 2 through 4 data, we obtained Cycle 1 observations of 14 stars that were known to 
 be variables or have $T_{\rm eff}$ and $\log$~g consistent with $\gamma$~Dor or $\delta$~Sct pulsations. 
 Nine of these stars were selected to lie within the ground-based hybrid temperature range 
 of 6900 K to 7350 K and have magnitudes between 14 and 15, while the other five stars 
 were variables selected from the All-Sky Automated Survey (ASAS; Pigulski et al. (2010))
 and had color indices consistent with being a $\delta$~Sct or $\gamma$~Dor star. 
 The $\delta$~Sct, $\gamma$~Dor, hybrid, and eclipsing binary stars from this 14 star sample 
 will be counted towards the total number of stars, but they will not be included in our 
 discussion of the instability strip location. 

The majority of the observations were made in long cadence ($\sim 30$ minute integrations per observation), 
although some early (Quarter 2 and 4) observations were made in 
short-cadence mode ($\sim 1$ minute integrations per observation). 
The bias towards long cadence data in our observations may affect our ability to determine the 
periods of $\delta$~Sct stars with frequencies above the Nyquist limit. 
However, the observations of 14 stars in Quarters 2 and 4 (Cycle 1) in both long and short cadence mode 
demonstrated that the $\delta$~Sct stars were easily detectable in long cadence mode 
(see also Balona and Dziembowski 2011). 
Also, \cite{murph13} show that it is possible to determine the correct super-Nyquist frequencies 
from long-cadence {\it Kepler} data. 
Once we determine that a star shows variability, the delta~Scuti nature can be confirmed.

\section{Frequency Analysis}

We analyzed the data from the 2768 stars in a consistent manner. 
The data were taken from the Mikulski Archive for Space Telescopes
(MAST)\footnote{For more information see http://archive.stsci.edu/kepler} 
database and we analyzed the PDCSAP\_FLUX corrected light curve files. 
These light curves have systematic error sources from the telescope and spacecraft removed 
\citep{KepArch}.
We first analyzed the light curves of the stars with tested software written by J.~Jackiewicz \citep{mcnam12}. 
This software processes the light curves by removing bad data points (infinities and points more than three 
sigma away from neighboring points), 
converts from {\it Kepler} flux (electrons/sec) to parts per million (ppm) for each light curve using the 
formula $f(t) = 10^6 ((F_K/y)-1)$ (where $y$ is the mean of the entire light curve or a low-order polynomial 
fit to the light curve, depending on artifacts that are present in the light curve),
performs a fast Fourier transform (FFT) on the light curve and plots the resulting power spectra 
on one page for ease of analysis. 
The results were visually inspected by at least two of the authors of this paper for variability. 
In addition, we analyzed the light curves with a locally created program that removed bad data 
points,  converted from {\it Kepler} flux (electrons/sec) to parts per million (ppm) for each light curve,
and performed a discrete FT  \citep{deeming75}. 
The results from the two methods were equivalent.
We analyzed each quarter of data separately as well as all the quarters in one data set. 
This allowed us to be sure the variations were indeed from our target star and not from 
another star sharing the same pixels in one orientation of {\it Kepler} (which we found in a few cases).
We did not perform any special stitching of data sets from one quarter to the next, since we
are only interested in determining whether a star was variable and what type of variable star it is.
We are less interested -- for now -- in the precise frequency or amplitude determinations, which would require
care in how the light curves from each quarter are "stitched" together.

Our analysis of the light curves and resultant FTs revealed that 1237 
of the 2768 stars did not show significant light variability at the 30~$\mu$mag (or 30~ppm) amplitude level. 
The KIC numbers of these stars are listed in Table 1. 
The remaining 1531 stars show light curve variability of some sort. 
We discuss the variable stars further in the following subsections by type.
We followed the criteria of Grigahc\'ene et al. (2010) and Uytterhoeven et al. (2011) in that we required stars to 
have at least three frequencies that are not obvious harmonics or combination peaks of each other 
(the exception is the HADS group; see below) to be classified as a
$\gamma$~Dor, $\delta$~Sct, or hybrid star candidate. 
We specifically excluded harmonic frequencies and frequencies below about 0.2 $ {\rm d}^{-1}$ from consideration.
The former peaks are not independent modes and periods over 5 days would almost certainly be due to rotation or
binary motion. 
To do this, we visually examined the light curve, the FT, and examined the frequencies of the larger peaks in the FT. 
We have further criteria for accepting a star as a hybrid candidate, as discussed in the next section. 
We generally followed Balona  et al. (2011) for the criteria for eclipsing binaries and rotating, spotted stars and these
criteria are similar to that of Uytterhoeven et al. (2011). 
We also used the classification scheme of Balona et al. (2011) wherever possible to make the comparison of our 
results to theirs easier.

\clearpage

\begin{deluxetable}{llllll}
\tabletypesize{\scriptsize}
\tablecaption{Nonvariable (amplitude $< 30$~ppm) stars found in our survey}
\tablewidth{0pt}
\tablehead{
\colhead{KIC {\#}} & \colhead{KIC {\#}} & \colhead{KIC {\#}} & \colhead{KIC {\#}} & \colhead{KIC {\#}} & \colhead{KIC {\#}} }
\startdata
   1026536 &    1433534 &    2974588 &    2991687 &    3117547 &    3233511 \\ 
   3236044 &    3245621 &    3329462 &    3341092 &    3342912 &    3343104 \\ 
   3343915 &    3353469 &    3356332 &    3429786 &    3439956 &    3444020 \\ 
   3444187 &    3444426 &    3454000 &    3457689 &    3457925 &    3553769 \\ 
   3557803 &    3558758 &    3640389 &    3643325 &    3646321 &    3729981 \\ 
  \enddata
\tablecomments{Table 1 is published in its entirety in the 
electronic edition of the {\it Astronomical Journal}.  A portion is 
shown here for guidance regarding its form and content.}
\end{deluxetable}

\section{Discussion of Variable Stars}

The Balona et al. (2011) study considered three types of $\gamma$~Dor/$\delta$~Sct variables. 
The first two types are $\gamma$~Dor and $\delta$~Sct stars. 
In addition, some stars show both types of pulsation behavior simultaneously and are called hybrid stars. 
We classify a star as hybrid if it satisfies the following three criteria (see Grigahc\'ene et al. (2010) and Uytterhoeven et al. (2011)): 
1) frequencies are detected both in the $\delta$~Sct ($> 5 {\rm d}^{-1}$ or $> 58 \mu$Hz) and gamma 
Doradus ($< 5 {\rm d}^{-1}$ or $< 58 \mu$Hz) frequency regimes; 
2) the amplitudes in the two regimes are roughly comparable (within a factor of 7); and 
3) at least two frequencies (that are not obvious harmonics or combination frequencies) are found in each 
regime with amplitudes greater than 40 ppm. 
The $T_{\rm eff}$ and $\log$~g range for these stars also encompasses some other objects that show 
light curve variability. 
Contact eclipsing binary stars, especially W Ursae Majoris (W~UMa) stars 
(contact binaries with both stars being typically early F spectral types) also fall into the hybrid star parameter space. 
We also found several detached eclipsing binaries and other types of binary star systems that will 
be discussed in more detail later. 

Stars with rotationally modulated starspots can also lie in the same part of the 
$T_{\rm eff}$ and $\log$~g range as $\delta$~Sct, $\gamma$~Dor, and hybrid stars. 
The spots can induce a near monotonic behavior or can show Òtraveling waveÓ patterns. 
The spots can also come and go in these stars, so the light curves can sometimes be
irregularly modulated. 
Since most main-sequence stars in the $\delta$~Sct/$\gamma$~Dor region are fairly rapid 
rotators ($50~{\rm km~s}^{-1} < v \sin i < 200~{\rm km~s}^{-1}$), we expect the rotation periods to be between 
0.5 and 2 days. 
Because the rotation frequency range
overlaps the $\gamma$~Dor pulsation frequency range, we require at least 
two frequencies (that are not obvious harmonics or combination frequencies) with sharp peaks 
before we can classify a star as being a $\gamma$~Dor candidate. 
Our experience is that many rotating stars tend to show an indistinct cluster of low frequency 
peaks (the ÒROTÓ category), although sometimes there are one (SPOTV) or two (SPOTM) fairly 
distinct peaks present; see the ÒRotationally Variable starsÓ section for 
a description. 
When two peaks are present in a SPOTM star, one peak is the harmonic (twice the frequency) of the first due to the 
non-sinusoidal pulse shape. 
The large number of possible mechanisms for light curve variability led us to 
create several categories of stars based on the morphology of the light curve,
and we borrowed several of the categories from Balona (2011). 
In some cases, the light curve morphology is indicative of specific physical behavior, 
but in other cases the physical behavior cannot be uniquely determined. 
We list the types of stars and the number found in Table 2. 
We discuss representative members of each group below. 

\begin{deluxetable}{lll}
\tabletypesize{\scriptsize}
\tablecaption{Morphological Classification of variable star types}
\tablewidth{0pt}
\tablehead{
\colhead{Category} & \colhead{Sub-category} & \colhead{number} }
\startdata
$\gamma$~Dor & Asymmetric (ASYM)         & 33 \\
          (207 stars)   & Symmetric (SYM)              & 88 \\
                                & multiple periods (MULT)  & 86 \\
$\delta$~Sct            & High amplitude (HADS)   & 47\\
         (84 stars)      & multiple periods (MULT) & 33 \\
                                 & "other"                                 & 4 \\
 Hybrid                   & $\gamma$ Dor dominant     &    7 \\
       (32 stars)        & $\delta$~Sct dominant        &    7 \\
                                & roughly equal                    & 18 \\
 Binary                   & EA (detached)                    & 17 \\
       (76 stars)        & EB (contact)                        & 52 \\
                                & "transit"                                &   4 \\
                                & "heartbeat"                          &  3 \\
 Rotation                & SPOTV (dominant period) &  75 \\
     (1132 stars)     & SPOTM (traveling wave)    & 109 \\
                                & ROT (dominant low frequency & 844\\
                                & VAR (low amplitude, type unknown) & 103 \\
\enddata
\end{deluxetable}

\subsection{$\gamma$~Dor Star Candidates} 

We detected 6 $\gamma$~Dor candidate stars in our limited sample of 14 stars from 
Quarters 2 and 4, and another 201 candidate $\gamma$~Dor stars in the larger sample, for 
a total of 207 stars. This is 13.4{\%} of the variable stars. 
The $\gamma$~Dor stars are identified in Table 3 by KIC number, {\it Kepler} magnitude $K_p$, $T_{\rm eff}$, 
$\log$~g, and category. 
The $T_{\rm eff}$ values have been adjusted upward by 200~K from the KIC catalog to account for the systematic 
temperature offset described by \cite{pin12}.
We have three categories, first described by \cite{bal11}. 
They are: ASYM, where the amplitude of the beat pattern above the mean level is considerably 
greater (about 2 times) than the amplitude below the mean level; 
the SYM category is similar to ASYM except that the excursions above and below the mean 
are nearly symmetric; and the MULT class shows a multitude of relatively low amplitude modes. 
We show three examples of our discoveries in Figure 3, with the light curve for a 
representative quarter in the left panel and the resultant FT in the right panel. 
KIC6128330 is a typical ASYM star whose maximum excursions about the mean are about 
three times the minimum excursions. 
KIC7191683 is a SYM star and the closely spaced frequencies in the FT result in a strong 
ÒbeatÓ pattern in the light curve. 
Finally, KIC6210324 is a MULT star that has numerous FT peaks between 0.25 and 5 c~$d^{-1}$. 
Our discoveries have frequency patterns ranging from simple patterns of a few modes 
to complicated FTs that indicate rich pulsators that will be promising for asteroseismology. 
Although the typical period range for $\gamma$~Dor pulsations is 0.33 to 3 days, 
we saw some stars with multi-periodic pulsations longer than 3 days that looked identical 
to shorter period stars that we identify as $\gamma$~Dor candidates. 
All of these stars were in our ÒMULTÓ category and none had periods longer than 4 days. 
While some of these candidates may indeed be $\gamma$~Dor stars, some may also 
have rotating spots present, either along with pulsations or instead of pulsations. 
These stars deserve further scrutiny. 

We plot our $\gamma$~Dor (plus $\delta$~Sct and hybrid) stars on a $T_{\rm eff}$, $\log$~g 
diagram in Figure 4. 
Most of our $\gamma$~Dor stars lie on the cool side of the ground-based instability strip, 
although many lie within the strip as well. 
Some $\gamma$~Dor stars are hotter than the $\gamma$~Dor instability region, 
but all of these lie within the $\delta$~Sct instability region. 
This behavior was also seen by Balona et al. (2011, their Figure 4) and Uytterhoeven et al. (2011, 
their Figure 10b). 
We examined the amplitude distribution of our discoveries and find that the majority of them 
have peak mode amplitudes of 100 to 500 ppm, although significant numbers are found 
with amplitudes up to $10^4$~ppm (0.01 mag).

\begin{figure}
\epsscale{.80}
\plotone{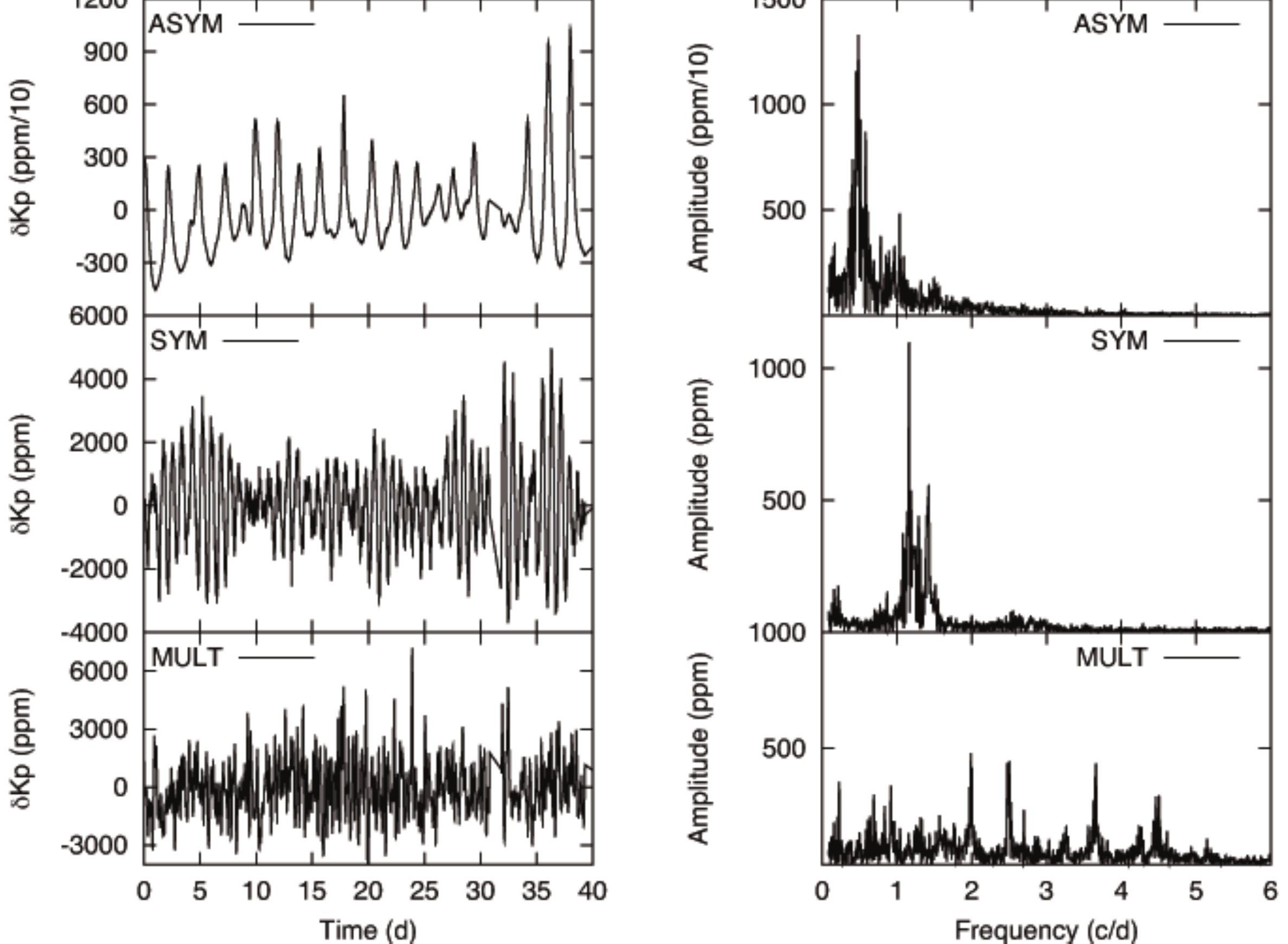}
\caption{Light curves (left panels) and Fourier Transforms (right panels) for three representative $\gamma$~Dor star
candidates. 
KIC6128330 (top row) is an example of an ASYM star, KIC7191683 (middle row) is a SYM star, and KIC6210324 
is an example of a MULT star. Note the different FT of the MULT star compared to the ASYM and SYM stars. 
\label{fig3}}
\end{figure}

\clearpage

\begin{figure}
\epsscale{.80}
\plotone{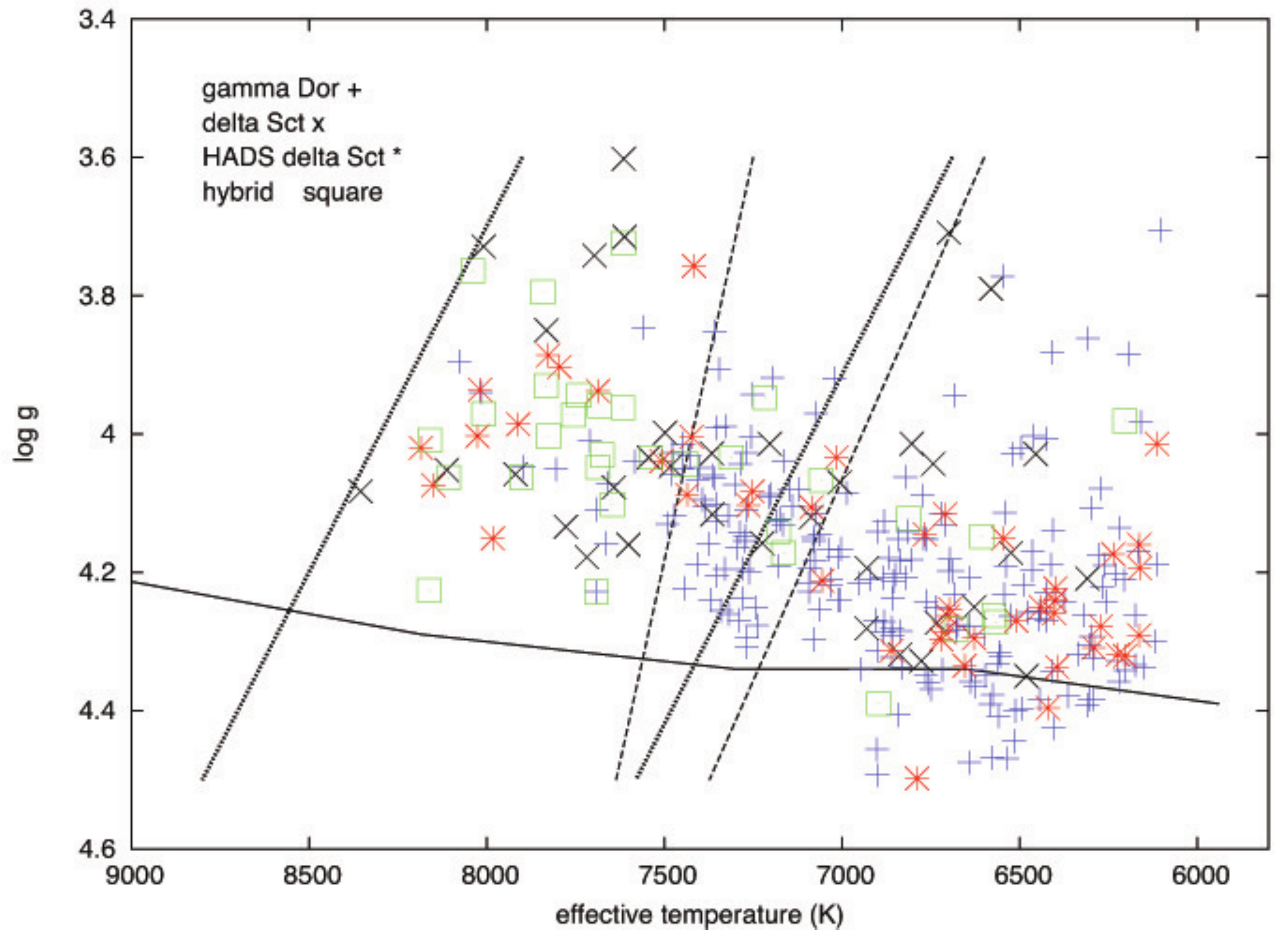}
\caption{Location of the $\gamma$~Dor, $\delta$~Sct, and hybrid star candidates from our sample in the 
$T_{\rm eff }$, $\log$~g diagram. 
The ground-based $\delta$~Sct (thick dotted lines) and $\gamma$~Dor (thin dashed lines) instability strips are indicated, 
along with the zero-age main sequence (solid line). 
The $\gamma$~Dor star candidates are blue  ``+'' signs, $\delta$~Sct star candidates 
are black  ``x'' signs, high-amplitude $\delta$~Sct (HADS) star candidates are red  ``*'' signs, 
and hybrid star candidates are green squares. There are no variable stars hotter than the $\delta$~Sct blue edge, 
and there are relatively few $\gamma$~Dor star candidates hotter than the $\gamma$~Dor blue edge. 
\label{fig4}}
\end{figure}

\clearpage

\begin{deluxetable}{lccclccc}
\tabletypesize{\scriptsize}
\tablecaption{$\gamma$~Dor star candidates found in our survey}
\tablewidth{0pt}
\tablehead{
\colhead{KIC {\#}} & \colhead{$K_p$} & \colhead{${\rm T}_{\rm eff}$} & \colhead{log g} & \colhead{class} &
\colhead{Freq. Range}&\colhead{Ampl. high}&\colhead{Freq. high}
}
\startdata
                   &          &           &             &            & (${\rm d}^{-1}$) & (ppm) & (${\rm d}^{-1}$) \\
  2167444 & 14.1 & 7140 & 4.1 & MULT &  $0.5 - 5.2$       &  1730  &    0.8082            \\
  2448307 & 14.0 & 7350 & 3.9 & MULT &  $0.2 - 3.5$        &  1390  &   1.2516            \\
  2579595 & 14.1 & 7150 & 4.1 & ASYM &  $0.7 - 1.7$       &   5175  &    1.2205           \\
  2581964 & 14.0 & 7410 & 4.2 & ASYM &  $0.4 - 1.5$       &   5914  &   0.5389            \\
  2857178 & 14.6 & 7440 & 4.2 & ASYM &  $0.9 - 2.0$       &     392  &   1.7313            \\
  2974858 & 14.4 & 7010 &  4.2 &  MULT& $0.75 - 3.1$    &       49 &    2.0287            \\
  \enddata
\tablecomments{Table 3 is published in its entirety in the 
electronic edition of the {\it Astronomical Journal}.  A portion is 
shown here for guidance regarding its form and content. ``Ampl. high'' and ``Freq. high'' refer to the amplitude and
frequency of the highest amplitude mode in the FT. The ${\rm T}_{\rm eff}$ and {$\log$~g} values are
rounded from the {\it Kepler} input catalog.}
\end{deluxetable}

\subsection{$\delta$~Sct Star Candidates} 

In our study, we find 81 candidate $\delta$~Sct stars (and 3 more candidates in our limited sample of 14 stars, 
for a total of 84 candidate stars, or 5.4{\%} of the variable stars). 
The $\delta$~Sct candidate stars are identified in Table 4 by KIC number, {\it Kepler} magnitude $K_p$, $T_{\rm eff}$, 
$\log$~g, and category. 
We use three classifications of $\delta$~Sct stars here. ÒHADSÓ stands for High Amplitude $\delta$~Sct Stars. 
Many of these are monoperiodic, but the short ($<$ 6 hour) period of the dominant peak implies that the
dominant peak cannot be binary orbital period or rotation period due to the impossibly high velocities (but see the discussion 
below).
The next class is ÒMULTÓ, which covers the objects with a rich spectrum of pulsation modes. 
Stars that do not fall in either of these two categories are called ``other''. 
We show three examples of our discoveries in Figure 5, with the light curve for a representative 
quarter in the left panel and the resultant FT in the right panel. 
KIC2581626 is a multiperiodic HADS whose closely spaced modes leads to strong ÒbeatingÓ in the light curve. 
Second, KIC 5707205 is a MULT star that has several bands between 8 and 18 c~$d^{-1}$. 
Finally, KIC 6304420 is in the ``other'' category. 
There are two dominant frequency bands that result in a strongly modulated light curve. 
Our discoveries have frequency patterns ranging from simple patterns of a few modes to 
complicated FTs that indicate rich pulsators that will be promising for asteroseismology. 

We plot our $\delta$~Sct star candidates on a $T_{\rm eff}$, $\log$~g diagram in Figure 4. 
Most of our $\delta$~Sct star candidates lie on the cool side of the ground-based $\delta$~Sct instability strip, 
although many lie within the strip as well. 
We found far more cool $\delta$~Sct star candidates than was seen by Balona and Dziembowski (2011, their Figure 1) 
and Uytterhoeven et al. (2011, their Figure 10b). 
Our HADS stars have many (40 of 47) stars that show a single dominant peak (typically between 4 and 6 c~$d^{-1}$),
along with a small peak at half the frequency of the dominant peak and at least one harmonic peak. 
If the small peak (at half the frequency of the dominant peak) is the rotation frequency of a spotted star, then the 
implied rotation velocities are 150 to 250 km~s$^{-1}$, which is physically plausible. 
We note that Balona (2011) also found that the dominant low frequency mode appears to be twice the rotation 
frequency. 
The remaining $\delta$~Sct star candidates would have a temperature distribution peaked between 6500 and 8000~K, 
which would be closer to that of Balona and Dziembowski (2011) and Uytterhoeven et al. (2011).
Spectroscopic observations of rotationally broadened absorption lines in our HADS candidates will be required to
confirm this hypothesis. 

We also examined the amplitude distribution of our discoveries and find a bimodal distribution. 
The multiperiodic stars comprise the entire sample with amplitudes below $10^4$~ppm (0.01 mag)
 and only one star has an amplitude greater than this. 
 By our definition, the HADS stars have amplitudes greater than 
 $10^4$~ppm (0.01 mag) and there are 42 of these (half of our discoveries), while 9 stars 
 have amplitudes greater than $10^5$~ppm (0.1 mag). 
 If we compare our amplitude distribution to Balona and Dziembowski (2011), we find a 
 dearth of pulsators with amplitudes below 1000 ppm. 
 It is possible that short cadence data might find additional low amplitude and/or short period $\delta$~Sct stars.
 We note, however, that our sample (see Figure 1) contains relatively few stars between 7000 and 8000 K
 below magnitude 14 (for our contamination factor of $<0.05$), suggesting that there just are not any more 
 hot stars available, probably because these stars are far enough away that they lie between spiral arms in our Galaxy
 (see discussion in Section~4.6). 

\begin{figure}
\epsscale{.80}
\plotone{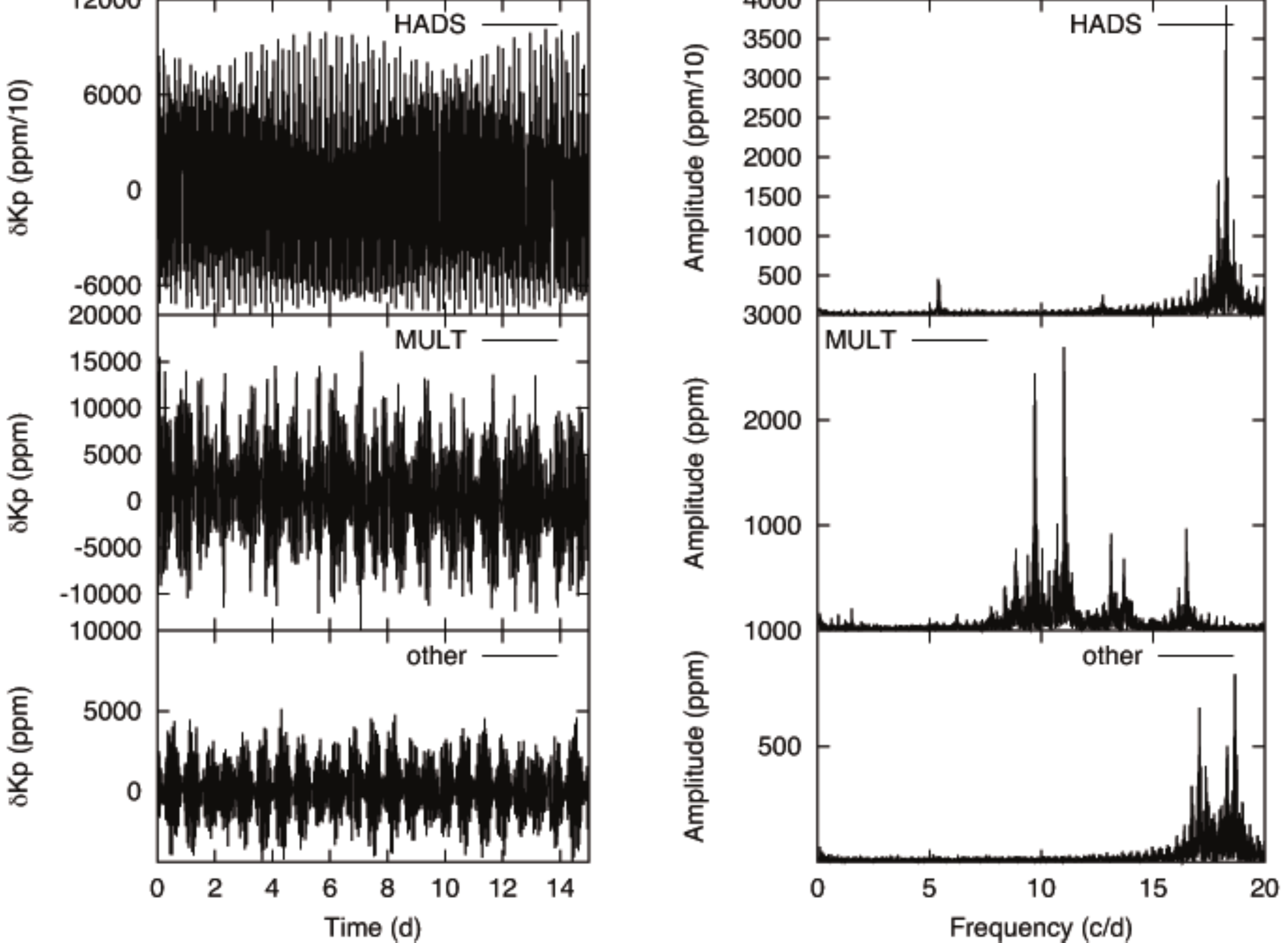}
\caption{Light curves (left panels) and Fourier Transforms (right panels) for three representative $\delta$~Sct star
candidates. 
KIC2581626 (top row) is an example of a HADS star, KIC5707205 (middle row) is an example of a MULT star, 
and KIC6304420 (bottom row) is an ``other'' star. 
\label{fig5}}
\end{figure}

\clearpage

\begin{deluxetable}{lccclccc}
\tabletypesize{\scriptsize}
\tablecaption{$\delta$~Sct star candidates found in our survey}
\tablewidth{0pt}
\tablehead{
\colhead{KIC {\#}} & \colhead{$K_p$} & \colhead{${\rm T}_{\rm eff}$} & \colhead{log g} & \colhead{class} &
\colhead{Freq. Range}&\colhead{Ampl. high}&\colhead{Freq. high}
}
\startdata
                   &          &           &             &            & (${\rm d}^{-1}$) & (ppm) & (${\rm d}^{-1}$) \\
   2581626 &  15.0 &   8030 &  4.0 &  HADS & $18.2 - 18.3$ & 18,090 & 18.2553\\
   2972514 &  14.0 &   6550 &  4.2 &  HADS & $3.95 - 4.05$ & 57,465 &    3.9993\\
   3119295 &  14.3 &   7440 &  4.1 &  HADS & $4.55 - 4.65$ & 49,230 &    4.5944\\
   3953144 &  14.7 &   8150 &  4.1 &  HADS & $8.0 - 18.5$    &  8680  &    9.9374\\
   4036687 &  15.2 &   6400 &  4.2 &  HADS & $6.7 - 6.8$      & 28,080 &   6.7302\\
   4066203 &  14.1 &   6700 &  4.3 &  HADS & $5.65 - 5.75$ & 14.250 &   5.7163\\
   4243668 &  14.7 &   6510 &  4.3 &  HADS & $5.2 - 5.3$     &  10,060 &   5.2646\\
   4374279 &  14.6 &   7060 &  4.2 &  HADS & $5.85 - 5.95$ &   4000  &   5.9070\\
   4466691 &  14.3 &   6710 &  4.3 &  HADS & $4.1 - 4.2$     &  28,980 &   4.1723\\
   4547067 &  14.7 &   6240 &  4.2 &  HADS & $4.9 - 5.0$     &  14,190 &    4.9662\\
   4651526 &  14.8 &   6440 &  4.2 &  HADS & $6.7 - 6.8$     &  22,055 &   6.7715\\
   4995588 &  15.0 &   7600 &  4.2 &  other   & $23.5 - 24.0$ &    4300 &  23.7488\\
   5027750 &  14.8 &   6930 &  4.2 &  MULT  & $10.0 - 24.0$ &    2670 & 11.8209\\
   5284701 &  14.4 &   7720 &  4.2 &  MULT  & $15.2 - 24.0$ &      760 & 19.3581\\
   5286485 &  14.8 &   6840 &  4.3 &  MULT  & $7.6 - 10.1$   &    4270 & 10.0890\\
   5353653 &  15.1 &   6710 &  4.1 &  HADS & $5.45 - 5.55$ &  11,205 &  5.4730\\
   5357882 &  14.7 &   6630 &  4.3 &  HADS & $5.15 - 5.25$ &  21,745 &  5.2004\\
   5534340 &  14.6 &   7250 &  4.1 &  HADS & $5.4 - 16.0$   &  12,280 &  6.0612\\
   5611763 &  14.9 &   6400 &  4.2 &  HADS & $5.55 - 5.65$ &  15,260 &  5.5868\\
   5707205 &  14.3 &   7220 &  4.2 &  MULT  & $6.0 - 17.0$   &    7890  &  9.7293\\
   5788165 &  14.4 &   8110 &  4.1 &  MULT  & $8.0 - 23.0$   &    3670  & 17.9811\\
   5966237 &  14.9 &   6790 &  4.5 &  HADS  & $5.05 - 5.15$ & 11,780 &   5.0919\\
   5978805 &  14.1 &   6740 &  4.0 &  other    & $21.2 - 24.0$ &      250 &  23.7209\\
   6271512 &  14.2 &   7830 &  3.9 &  HADS  & $10.3 - 10.4$ &    8300 & 10.3721\\
   6304420 &  14.4 &   7480 &  4.0 &  MULT   & $16.9 - 18.6$ &    1500 & 18.5395\\
   6344429 &  14.7 &   7020 &  4.0 &  HADS  & $4.85 - 4.95$ &  12,600 &   4.8972\\
   6442207 &  15.5 &   7980 &  4.2 &  HADS  & $4.35 - 4.45$ &   11,000 &  4.4000\\
   6444630 &  14.6 &   6160 &  4.2 &  HADS  & $6.0 - 6.1$     &      4200 &   6.0558\\
   6672071 &  14.9 &   6200 &  4.3 &  HADS  & $ 5.65 - 5.75$&  100,000&  5.7070\\
   6696050 &  14.3 &   7830 &  3.8 &  MULT  & $12.0 - 12.1$  &      2000 & 12.028\\
   6778487 &  14.8 &   6420 &  4.4 &  HADS  & $5.05 - 5.15$ &    91,000 &  5.095\\
   6836820 &  14.5 &   6270 &  4.3 &  HADS  & $4.4 - 4.5$     &  110,000 &  4.4720\\
   6870432 &  14.0 &   7090 &  4.1 &  MULT  & $20.7 - 20.8$ &        1500 & 20.772\\
   6955650 &  14.2 &   7510 &  4.0 &  HADS  & $3.75 - 3.85$ &       4100 &   3.7967\\
   7048016 &  14.0 &   6480 &  4.4 &  MULT  & $16.0 - 16.1$ &       5470 & 16.0597\\
   7124161 &  14.7 &   7690 &  3.9 &  HADS  & $4.65 - 4.75$ &    25,000 &   4.6869\\
   7347529 &  14.1 &   7610 &  3.6 &  MULT  & $11.2 - 11.8$ & 654,900 &  11.7416\\
   7381616 &  14.5 &   6930 &  4.3 &  MULT  & $8.8 - 22.0$   &    3800    &  15.014\\
   7521682 &  14.8 &   6660 &  4.3 &  HADS  & $ 5.55 - 5.65$&   7120    &  5.609  \\
   7601767 &  14.5 &   6770 &  4.1 &  HADS  & $4.05 - 4.15$ &  88,600 &  4.1121\\
   7617649 &  14.6 &   6160 &  4.3 &  HADS  & $5.4 - 5.5$     & 175,000 &  5.433\\
   7668283 &  14.6 &   7420 &  3.8 &  HADS  & $3.65 - 3.75$ &  12,500 & 3.7143\\
   7750215 &  14.3 &   8360 &  4.1 &  MULT  & $15.2 - 23.0$ &      5795 & 16.4151\\
   7905603 &  14.3 &   7500 &  4.0 &  MULT  & $7.9 - 18.5$   &      9235 & 18.233\\
   7937097 &  14.3 &   6220 &  4.3 &  HADS  & $4.65 - 4.75$ &     7300 & 4.6869\\
   7948091 &  14.8 &   6290 &  4.3 &  HADS  & $6.5 - 6.6$     &      3330 & 6.526\\
   7984934 &  14.1 &   6160 &  4.2 &  HADS  & $6.8 - 6.9$     &   19,500 & 6.874\\
   8052082 &  14.6 &   6780 &  4.3 &  MULT  & $15.9 - 24.0$ &      1145 & 18.3534\\
   8087649 &  14.4 &   7260 &  4.1 &  HADS  & $4.1 - 4.2$    &       4700 & 4.1682 \\
   8090059 &  14.0 &   8020 &  3.9 &  HADS  & $18.3 - 18.4$ &     4750 & 18.3534\\
   8144212 &  14.2 &   7360 &  4.1 &  MULT  & $9.3 - 9.4$    &       1600 &   9.349\\
   8150307 &  14.8 &   7440 &  4.1 &  HADS  & $14.65-14.75$&   4500 & 14.6977\\
   8245366 &  11.2 &    N.A.  & N.A.&  MULT  & $6.9 - 12.2$  &  808,370& 11.9743\\
   8248296 &  14.1 &   7920 &  4.1 &  MULT  & $10.4 - 22.8$ &        465 & 13.556\\
   8248967 &  15.1 &   6720 &  4.3 &  HADS  & $3.3 - 3.4$     &    15,000 & 3.3715\\
   8249829 &  14.3 &   7090 &  4.1 &  HADS  & $16.7 - 16.8$ &      2450 & 16.7814\\
   8315263 &  14.3 &   7700 &  3.7 &  MULT  & $13.0 - 18.8$ &      7300 &  13.2372\\
   8322016 &  14.1 &   6800 &  4.0 &  other    & $14.4 - 24.5$ &       200 &  24.0372\\
   8323981 &  14.3 &   8010 &  3.7 &  MULT  & $8.0 -23.2$    &       120 &  20.4372\\
   8393922 &  14.0 &   7540 &  4.0 &  MULT  & $9.8 - 22.0$   &      2000 & 13.0791\\
   8508096 &  14.2 &   7420 &  4.0 &  HADS  & $22.1 - 22.2$ &     1500 & 22.1209\\
   8516900 &  14.2 &   6580 &  3.8 &  MULT  & $5.0 - 20.0$   &     2600 & 16.3628\\
   8648251 &  14.6 &   6520 &  4.2 &  MULT  & $15.1 - 19.6$ &    5240 & 15.1256\\
   8649814 &  14.4 &   6700 &  3.7 &  MULT  & $7.6 - 15.6$   &    5685 &    7.600 \\
   8960514 &  14.4 &   6730 &  4.3 &  MULT  & $6.65 - 6.75$ &    3100 &   6.6977\\
   8963394 &  14.5 &   6110 &  4.0 &  HADS  & $5.3 - 5.4$     &   11,840&  5.3581\\
   9075949 &  14.6 &   6400 &  4.3 &  HADS  & $5.65 - 5.75$ &   10,930&  5.6837\\
   9077483 &  15.4 &   6400 &  4.3 &  HADS  & $5.4 - 5.5$      &  25,155 & 5.4276\\
   9137819 &  15.0 &   7800 &  3.9 &  HADS  & $3.6 - 3.7$      &  17,410   & 3.6595\\
   9202969 &  14.0 &   6700 &  4.3 &  HADS  & $4.95 - 5.05$ &     4310 &   4.9810\\
   9214444 &  14.2 &   7610 &  3.7 &  MULT  & $8.1 - 24.0$    &      540 &  21.6372\\
   9364179 &  14.1 &   7200 &  4.0 &  MULT   & $18.0 - 20.0$  &    1235 &  18.200\\
   9594857 &  11.0 &    N.A.  &  N.A.&  MULT   &$5.0 - 10.0$     &  862,070& 9.9695\\
   9613175 &  14.2 &   7370 &  4.0 &  MULT   & $18.7 - 24.0$  &         330 & 20.6741\\
   9613575 &  14.4 &   6310 &  4.2 &  MULT   & $8.5 - 24.0$    &       3360 & 13.1628\\
   9614153 &  14.2 &   6460 &  4.0 &  MULT   & $10.0 - 16.4$  &       2270 & 13.2186\\
   9700322 &  12.7 &    N.A.  & N.A.&  HADS  & $9.6 - 12.6$    &  392,840 & 12.6259\\
   9706609 &  14.1 &   7640 &  4.1 &  MULT+rot& $5.0 - 9.0$  &          420 &   7.3767\\
   9724292 &  14.4 &   7010 &  4.1 &  MULT  & $10.8 - 16.4$  &        2810 &13.8000\\
   9942562 &  14.6 &   6630 &  4.2 &  other    & $6.75 - 6.85$ &           550 &  6.8000 \\
  10451090 &   9.2 &   7780 &  4.1 &  MULT  & $10.0 - 20.0$ &      31,050 & 10.6695\\
  11143576 &  15.1 &  8180 &  4.0 &  HADS & $15.7 - 15.8$ &        6480 & 15.7719\\
  11704101 &  15.4 &  6860 &  4.3 &  HADS  & $5.8 - 5.9$    &  959,070  &   5.8527\\
  11852985 &  14.4 &  7910 &  4.0 &  HADS  & $19.85-19.95$& 10,455 &  19.8915\\
  \enddata
\tablecomments{``Ampl. high'' and ``Freq. high'' refer to the amplitude and
frequency of the highest amplitude mode in the FT. The ${\rm T}_{\rm eff}$ and {$\log$~g} values are
rounded from the {\it Kepler} input catalog.}
\end{deluxetable}

\clearpage
 
\subsection{Hybrid Star Candidates} 

We discovered one hybrid star candidate in our limited sample of 14 objects and only 31 hybrid
candidates in our larger sample. Thus, the hybrid stars constitute 2.1{\%} of the variable star candidates.
We use three classifications: $\gamma$~Dor-dominated FTs, $\delta$~Sct-dominated FTs, and 
ones where the $\gamma$~Dor and $\delta$~Sct amplitudes are within a factor of three of each other
and are placed in the ``equal'' bin. 
The hybrid star candidates are identified in Table 5 by KIC number, {\it Kepler} magnitude 
$K_p$, $T_{\rm eff}$, $\log$~g, and category. 
We show three of the hybrid stars from our limited data sample in Figure 6, in two columns. 
KIC5561007 is representative of our $\gamma$~Dor-dominated hybrid star candidates, 
while KIC3657237 is one of the few $\delta$~Sct-dominated hybrid star candidates.
KIC2855026 has prominent peaks from 1 to 20 c~$d^{-1}$ and is in the``equal'' amplitude category.
We plot our hybrid star candidate discoveries on a $T_{\rm eff}$, $\log$~g diagram in Figure 4. 
Most of our hybrid star candidates lie within the ground-based $\delta$~Sct instability strip, although 
one-third lie on the cool side of the instability strip. 
This is in marked contrast to the $\gamma$~Dor and $\delta$~Sct stars. 
This behavior was also seen by Uytterhoeven et al. (2011, Figure 10b). 
In addition, the amplitude distribution of our hybrid stars shows that almost all have peak 
mode amplitudes below 3000 ppm in both the $\delta$~Sct and $\gamma$~Dor range. 
The relatively low mode amplitude works against our being able to detect faint 
(fainter than magnitude 15) hybrid stars. 

We note that Bouabid et al. (2013) showed that some stars whose frequency distribution would lead
to a hybrid classification could in fact be rapidly rotating $\gamma$~Dor stars whose g-modes have
been shifted to higher frequencies. 
Multi-color photometry and/or spectroscopic observations would be needed to determine the $p-$ or
$g-$mode nature of the modes and confirm whether a hybrid star candidate of ours is truly a hybrid star. 

\begin{figure}
\epsscale{.80}
\plotone{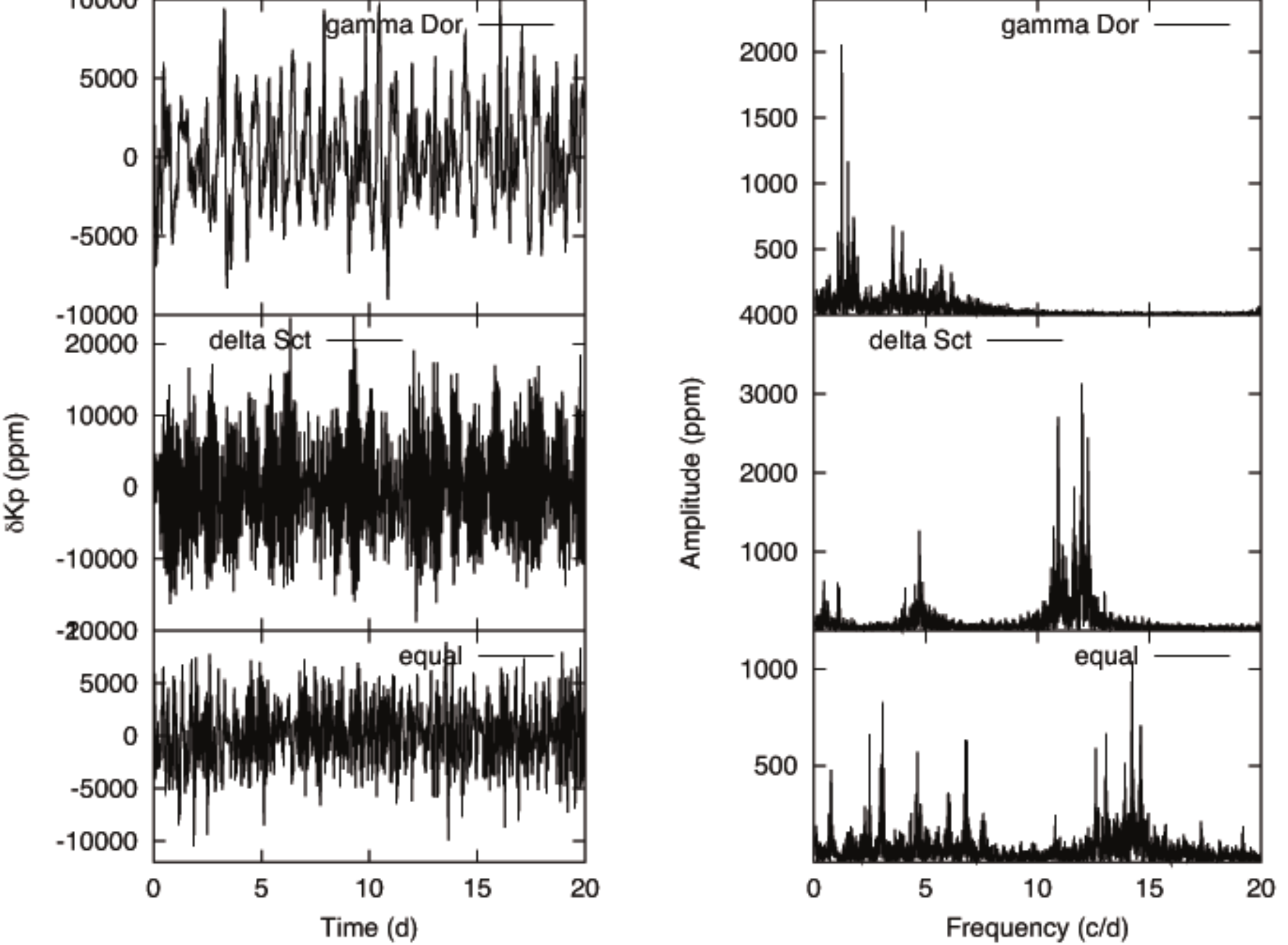}
\caption{Light curves (left panels) and Fourier Transforms (right panels) for three representative hybrid star
candidates. 
KIC5561007 (top row) is a $\gamma$~Dor dominant star, 
KIC3657237 (middle row) is a $\delta$~Sct dominant star, and KIC2855026 (bottom row) is an example of a star 
with near equal amplitudes in the $\gamma$~Dor and $\delta$~Sct ranges.  
\label{fig6}}
\end{figure}

\clearpage

\begin{deluxetable}{lccclcccc}
\tabletypesize{\scriptsize}
\tablecaption{Hybrid star candidates found in our survey}
\tablewidth{0pt}
\tablehead{
\colhead{KIC {\#}} & \colhead{$K_p$} & \colhead{${\rm T}_{\rm eff}$} & \colhead{log g} & \colhead{class} &
\colhead{$\gamma$~Dor Ampl.}&\colhead{$\gamma$~Dor Freq.}&\colhead{$\delta$~Sct Ampl.}&\colhead{$\delta$~Sct Freq.}
}
\startdata
                    &          &           &             &            &(ppm)   & (${\rm d}^{-1}$) & (ppm) & (${\rm d}^{-1}$) \\
   2301163 &  14.3 &   7540 &  4.0 &$\gamma$~Dor&  8175 & 2.1076  &  8925 & 19.7405\\
   2855026 &  14.1 &   7690 &  4.0 &                   equal&   855  & 3.0374  &  1665 & 17.3012\\
   3456100 &  14.1 &   8040 &  3.8 &$\delta$~Sct &    168  & 1.5304  &  1320 & 17.2411\\
   3657237 &  14.3 &   6570 &  4.3 &$\delta$~Sct &    600  & 1.0312  &  4585 & 11.9070\\
   3941524 &  14.1 &   7680 &  4.0 &             equal&     45   & 1.0537  &    180 & 23.6279\\
   4668676 &  14.1 &   6900 &  4.4 &$\delta$~Sct &   1900 & 1.9743 &  6865 & 15.3581\\
   4999763 &  15.0 &   7760 &  4.0 &              equal&   1295 & 1.3190 &    560 & 19.7767\\
   4999789 &  15.0 &   7650 &  4.1 &              equal&     940 &  1.6495 &  2720 & 15.4884\\
   5018191 &  14.5 &   7160 &  4.2 &$\delta$~Sct &     225 &  0.5238 &  2055 & 10.4465\\
   5466537 &  10.3 &   7180 &  4.1 &$\delta$~Sct &     205 &  2.7857 &  1080 & 22.2605\\
   5553489 &  14.4 &   6680 &  4.3 &$\gamma$~Dor&     150 &  1.0631 &      45 &    5.1238\\
   5561007 &  14.2 &   7690 &  4.2 &$\gamma$~Dor&   2075 &  0.9766 &    645 &   5.6465\\
   5771101 &  15.0 &   7440 &  4.0 &              equal&     720 &  0.7126 &  1135 &   7.8512\\
   5809732 &  14.1 &   6810 &  4.1 &              equal&     125 &  1.7827 &      80 &    8.4279\\
   5966212 &  15.1 &   7680 &  4.0 &              equal&    305 &  1.9121 &     800 & 15.8326\\
   6130261 &  14.9 &   7830 &  3.9 &$\gamma$~Dor&  3455 & 0.9930 &     590 &  21.7674\\
   6290877 &  14.1 &   8010 &  4.0 &  equal&     395 & 2.1051 &      885 & 16.7163\\
   6460258 &  14.2 &   7060 &  4.1 &$\gamma$~Dor&   2770 & 2.9810 &      365 &  5.1429\\
   6467349 &  14.6 &   6610 &  4.1 &              equal&     690 & 1.6005 &    1200 & 18.7814\\
   6586020 &  15.1 &   7310 &  4.0 &              equal&     425 & 3.2103 &      810 & 17.5814\\
   6936178 &  14.4 &   8160 &  4.0 &              equal&       45 & 1.8972 &       100 & 17.6930\\
   6960727 &  14.1 &   8160 &  4.2 &$\delta$~Sct &      125 & 2.1168 &     600 & 18.8000\\
   6974847 &  14.5 &   7840 &  3.8 &$\delta$~Sct &      500 & 1.4159 &   1825 & 17.9256\\
   7300263 &  14.5 &   6200 &  4.0 &              equal&        75 & 2.0958 &     140 &  17.8419\\
   7302192 &  14.5 &   6570 &  4.3 &              equal&      275 & 1.5818 &     725 & 15.8233\\
   7354531 &  14.5 &   7620 &  3.7 &              equal&    1440 & 1.2406 &   1890 & 13.0419\\
   7750216 &  14.0 &   8110 &  4.1 &$\delta$~Sct &       340 & 0.6519 &     750 & 22.2140\\
   8314246 &  14.0 &   7220 &  3.9 &              equal&     2350 & 1.8037 &  2600 & 14.1302\\
   9005210 &  14.0 &   7740 &  3.9 &              equal&       625 & 1.4650 &  1200 & 14.5953\\
   9402020 &  15.1 &   7620 &  4.0 &$\gamma$~Dor&       290 & 2.1449 &     170 & 19.4977\\
   9529640 &  14.5 &   7830 &  4.0 &              equal&      355 & 2.3575 &      700 & 21.6000\\
  10134571 &  14.6 &  7900 &  4.1 &             equal&       80 & 2.4418 &      190 & 21.6279\\  
  \enddata
\tablecomments{``$\gamma$~Dor Ampl.'' and ``$\gamma$~Dor Freq.'' refer to the amplitude and
frequency of the highest amplitude mode in the $\gamma$~Dor region ($<5$ ${\rm d}^{-1}$) of the FT, 
while ``$\delta$~Sct Ampl.'' and ``$\delta$~Sct Freq.'' refer to the amplitude and
frequency of the highest amplitude mode in the $\delta$~Sct region ($>5$ ${\rm d}^{-1}$) of the FT.
The ${\rm T}_{\rm eff}$ and {$\log$~g} values are rounded from the {\it Kepler} input catalog.}
\end{deluxetable}

\clearpage

\subsection{Binary Stars} 

We find 73 binary stars in our sample (and 3 more in our 14 star sample from Q2 and Q4), which is  4.9{\%} 
of our variable star sample. 
The binary stars are identified in Table 6 by KIC number, {\it Kepler} magnitude $K_p$, $T_{\rm eff}$, 
$\log$~g,  category, and orbital period (in days). 
Some of the stars show clear eclipses, although many show ellipsoidal variations that indicate
 two distorted stars seen from different positions as they orbit each other, to partially eclipsing 
distorted stars. 
We classified these stars into four categories (Balona 2011), with unphased light curve examples shown in Figure 7. 
The ``EA'' stars are detached binaries that show obvious eclipses, but no ellipsoidal effects, 
as shown by KIC8690001 in panel 1, which also obviously has an eccentric orbit. 
The orbital periods range from 1.3 to 24.1 days. 
The``EB'' stars are contact (or ellipsoidal) eclipsing binaries, such as KIC2161023 in panel 2. 
Many of them have periods of less than 1 day, making them likely W~UMa stars. 
The longest period for any of these is 3.72 days, so they are clearly very tight binaries 
(semimajor axis $< 10^7$ km). 
These are binary stars in eccentric  orbits where the stars become 
tidally distorted near periastron passage, causing a sudden rise in the light curve. 
The three stars have periods of 9.5, 12, and 23 days. KIC6963717 is the heartbeat star illustrated in the 
third panel. 
There are four stars labeled ``transit'' that have irregular light curves (due to starspots or 
rotation) punctuated by narrow eclipses, as shown by KIC7599004 in the bottom panel. 
The periods between the minima range from 4.9 hr to 5.5 d. 
We are not claiming that they are transits by dark objects, but we believe they are binary 
objects of some sort. 
We compared our list of binary stars to the {\it Kepler} binary star database (current as of 3/31/2014) and found that 44 
of our stars are not in this list. 
We highlight the new discoveries in boldface text in Table 6.
To the extent that the KIC photometric  $T_{\rm eff}$ and $\log$~g values have meaning for
binary stars, they are uniformly distributed in $T_{\rm eff}$, while the $\log$~g 
value increases with decreasing $T_{\rm eff}$. 
We refer the reader to Gaulme and Guzik (2014), who performed a more detailed analysis of these stars.

\begin{figure}
\epsscale{.80}
\plotone{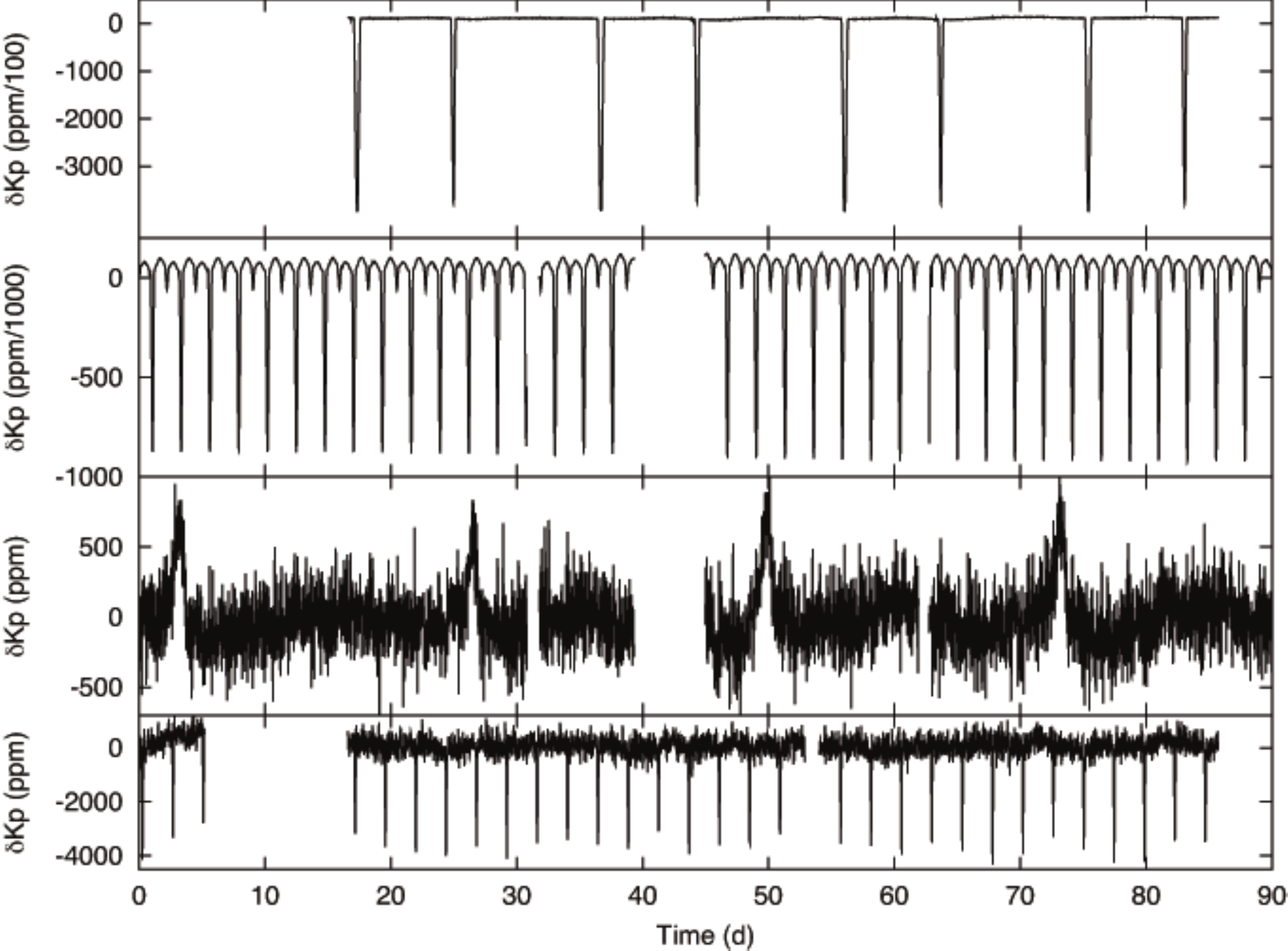}
\caption{Unphased light curves for four representative binary stars. KIC8690001 (top row) is an example of an EA 
(detached binary) star, KIC2161623 (second row) is an EB (ellipsoidal eclipsing binary) star,  KIC6963171 is a ``heartbeat'' star with tidal distortion near periastron passage causing 
the sharp rises at 3, 27, 50, and 73 days, and KIC7599004 is an example of a ``transit'' star with low amplitude 
eclipses.  \label{fig7}}
\end{figure}

%

\clearpage

\begin{deluxetable}{lcccll}
\tabletypesize{\scriptsize}
\tablecaption{Binary stars found in our survey}
\tablewidth{0pt}
\tablehead{
\colhead{KIC {\#}} & \colhead{$K_p$} & \colhead{${\rm T}_{\rm eff}$} & \colhead{log g} & \colhead{class} 
&\colhead{Per (d)}}
\startdata
   1433410         &  14.1 &   6200 &  4.0 &  EB           &0.283\\
   2161623         &  14.3 &   6710 &  4.1 &  EB           &2.283\\
   2449084         &  15.0 &   6380 &  4.4 &  EB           &0.740\\
   {\bf 2719436} &  14.0 &   7240 &  4.1 &  EB           &0.740\\
   2988984         &  15.3 &   7360 &  4.1 &  EB           &0.713\\
   3547111         &  14.4 &   6600 &  4.3 &  EB           &1.208\\
   3633901         &  14.9 &   7380 &  4.2 &  EB           &0.807\\
   4160006         &  14.3 &   6120 &  4.2 &  EA           &2.037\\
   4470124         &  14.9 &   6460 &  4.4 &  heartbeat&11.433\\
   4554004         &  15.1 &   6410 &  4.2 &  EB            &0.743\\
   4570326         &     9.8 &   N.A.  & N.A.&  EB            &1.122\\
   4739791         &  14.7 &   7740 &  3.9 &  EA+pulse&0.899\\
   4815612         &  15.2 &   6590 &  4.3 &  EA             &3.857\\
   4936680         &  14.3 &   8100 &  4.0 &  EB             &0.666\\
   5022908         &  14.3 &   6900 &  4.1 &  EB             &0.637\\
   5036966         &  14.3 &   6160 &  4.1 &  EB             &62.735\\
   5290305         &  14.3 &   6740 &  4.3 &  EB             &0.621\\
   5357682         &  14.6 &   6640 &  4.2 &  EB             &0.718\\
   5358200         &  15.0 &   7230 &  4.0 &  EB             &0.536\\
   5461570         &  14.8 &   7330 &  4.0 &  EB             &0.508\\
   {\bf 5524325} &  15.0 &   7380 &  4.0 &  EB             &0.626\\
   5606644         &  14.8 &   7740 &  4.2 &  EA             &0.862\\
   5615815         &  14.8 &   7330 &  4.1 &  EB+pulse&0.674\\
   {\bf 5616194} &  15.0 &   6580 &  4.2 &  EA             &0.623\\
   5858519         &  14.7 &   6130 &  4.3 &  EA             &4.182\\
   5878081         &  14.4 &   7900 &  3.8 &  EB+pulse&0.591\\
   {\bf 5962514} &  14.8 &   6770 &  4.3 &  EB            &1.585\\
   {\bf 6048106} &  14.1 &   6980 &  4.2 &  EB            &1.556\\
   {\bf 6220497} &  14.7 &   7450 &  3.9 &  EB            &1.332\\
   {\bf 6224853} &  14.4 &   7540 &  4.1 &  EB            &0.535\\
   {\bf 6948815} &  15.3 &   7630 &  4.0 &  EB            &1.556\\
   {\bf 6963171} &  14.1 &   6130 &  3.9 &  heartbeat&23.3\\
   7025851         &  14.4 &   6250 &  4.3 &  EA            &4.681\\
   {\bf 7107567} &  14.2 &   7100 &  4.2 &  transit       &0.809\\
   {\bf 7108433} &  15.1 &   7410 &  4.1 &  EB            &1.527\\
   {\bf 7365447} &  14.3 &   6800 &  4.3 &  EA            &2.471\\
   {\bf 7377343} &  14.4 &   6160 &  4.3 &  EA            &8.40\\
   {\bf 7436177} &  14.6 &   6270 &  4.3 &  EA            &10.50\\
   {\bf 7599004} &  14.8 &   6320 &  4.2 &  transit      &2.40\\
   {\bf 7700578} &  14.2 &   6890 &  4.1 &  EB            &1.505\\
   7740302         &  12.0 &   N.A.   & N.A.&  EB            &1.154\\
   {\bf 8153568} &  15.1 &   7000 &  4.1 &  EA            &3.652\\
   {\bf 8182360} &  15.3 &   7100 &  4.1 &  EB            &0.697\\
   {\bf 8183540} &  14.3 &   6630 &  4.2 &  EB            &0.343\\
   {\bf 8240861} &  15.3 &   6290 &  4.2 &  EB            &0.901\\
   8294484         &  14.7 &   6450 &  4.3 &  EB+spot  &1.013\\
   {\bf 8380743} &  14.0 &   6170 &  4.0 &  EB            &2.024\\
   {\bf 8455359} &  14.2 &   6840 &  4.0 &  EB            &2.959\\
   {\bf 8565912} &  14.7 &   6970 &  4.1 &  EB            &1.012\\
   {\bf 8579812} &  14.8 &   6720 &  4.1 &  EB            &0.658\\
   {\bf 8587078} &  14.0 &   6240 &  4.0 &  EB            &0.583\\
   {\bf 8690001} &  14.3 &   6140 &  3.9 &  EA            &19.2\\
   {\bf 8696327} &  14.6 &   6920 &  4.1 &  EB            &0.875\\
   {\bf 8736072} &  14.9 &   8030 &  4.1 &  EB            &0.477\\
   {\bf 8822555} &  14.4 &   6580 &  4.3 &  EB            &0.852\\
   {\bf 8895509} &  14.2 &   6260 &  3.8 &  heartbeat&9.767\\
   {\bf 8904714} &  14.8 &   6120 &  4.1 &  transit      &5.25\\
   {\bf 9101400} &  14.7 &   6610 &  4.3 &  EB            &1.647\\
   {\bf 9108058} &  14.3 &   6760 &  4.1 &  EB            &2.176\\
   {\bf 9205993} &  14.9 &   7320 &  4.1 &  EB            &1.226\\
   {\bf 9282687} &  14.1 &   6820 &  4.1 &  EB            &1.680\\
   {\bf 9291368} &  14.0 &   8090 &  3.7 &  EB            &3.717\\
   {\bf 9343862} &  15.0 &   7910 &  4.0 &  EB            &1.120\\
   {\bf 9479460} &  14.7 &   7770 &  3.7 &  EB            &2.089\\
   9514070         &  15.2 &   6790 &  4.1 &  EB            &0.607\\
   9658118         &  14.2 &   6420 &  4.3 &  EA            &24.06\\
   {\bf 9767392} &  14.7 &   6600 &  4.2 &  EA            &1.462\\
   {\bf 9832545} &  15.2 &   8100 &  3.9 &  EB            &1.012\\
   {\bf 9843435} &  14.8 &   7490 &  4.1 &  EB            &1.680\\
   {\bf 9899345} &  15.0 &   6800 &  4.1 &  transit      &1.333\\
   9936698         &  14.0 &   6590 &  4.2 &  EA            &5.712\\
   9954225         &  14.6 &   6290 &  4.2 &  EB            &1.324\\
 10141087         &  15.2 &   6620 &  4.2 &  EB           &0.469\\
 {\bf 11401845} &  14.4 &   7790 &  3.9 &  EA           &2.161\\
 11819135         &  15.1 &   6900 &  4.1 &  EB+pulse&1.902\\
 11867071         &  14.3 &   6600 &  4.4 &  EA+pulse?&2.964\\ 
  \enddata
\tablecomments{KIC numbers in {\bf boldface} are binary systems discovered 
in this study. The ${\rm T}_{\rm eff}$ and {$\log$~g} values are
rounded from the {\it Kepler} input catalog.}
\end{deluxetable}

\clearpage

\subsection{Rotating Stars} 

This class has stars that show starspots moving with rotation, and other sources of long-term 
frequency modulation. 
The rotating stars are identified in Table 7 by KIC number, {\it Kepler} magnitude $K_p$, $T_{\rm eff}$, 
$\log$~g, and category. 
It is the single biggest class of variable stars in our sample, with 1132 members, or 74.0{\%} of the variable stars
we found.
The large number of rotationally modulated stars is a reflection of the cooler stars in our sample.
We show three representative examples in Figure 8.  
The SPOTM category (represented by KIC3545661) shows a clear beat pattern (the light curve 
in Figure 8 is not long enough to fully show the beat cycle) and two dominant peaks in the FT. 
The modulated light curves are likely due to multiple spots rotating in and out of view. 
The SPOTV category (represented by KIC4276984) has a single dominant peak in the FT and shows a repeating light curve 
that can be explained by a single spot rotating in and out of view. 
The ROT category shows low frequency (typically $<1$ c~$d^{-1}$) power in the FT and a modulated 
light curve (KIC3248536 is an example), but no clear peaks in the FT like the SPOTM and 
SPOTV stars. 
Their variations may be due to rotation, but if so, the spots are not stable for more than a 
few rotation periods. 
Finally there are a number of stars that show low frequency variability, but with no clear period in
the light curve or the FT. 
We labeled such stars as ÒVARÓ, but we cannot ascribe an obvious physical mechanism 
to the variations. 
We show the distribution of our rotating stars in a $T_{\rm eff}$, $\log$~g diagram in Figure 9. 
As seen in Figure~9, almost all of these objects are cooler than the red edge of the $\gamma$~Dor 
instability strip.
The cool effective temperatures are consistent with rotation, since this implies the presence of deep convection zones
that would produce strong magnetic activity cycles. 
We note that a few stars are within (or are even hotter) than the $\delta$~Sct instability strip. 
Balona (2013) discusses instances of such stars in his sample. 
Our sample of nine stars with clearly identifiable peaks has properties similar to those of 
Balona (2013) in that two of the nine (22{\%}) have the harmonic peak as the highest 
amplitude (Balona 2013 has 25{\%}). 
Eight of our nine stars have implied rotational velocities (based on the fundamental 
mode frequency and KIC radius values) between 20 and 265 km/s. 
One star has a rotation velocity of 1.8 km/s; this could be a horizontal branch star. 
Given our small number of stars, the distribution is similar to Figure 8 in Balona (2013). 
Thus, our nine stars are also likely to be A-type stars with some sort of rotating feature, 
such as starspots.  

\begin{figure}
\epsscale{.80}
\plotone{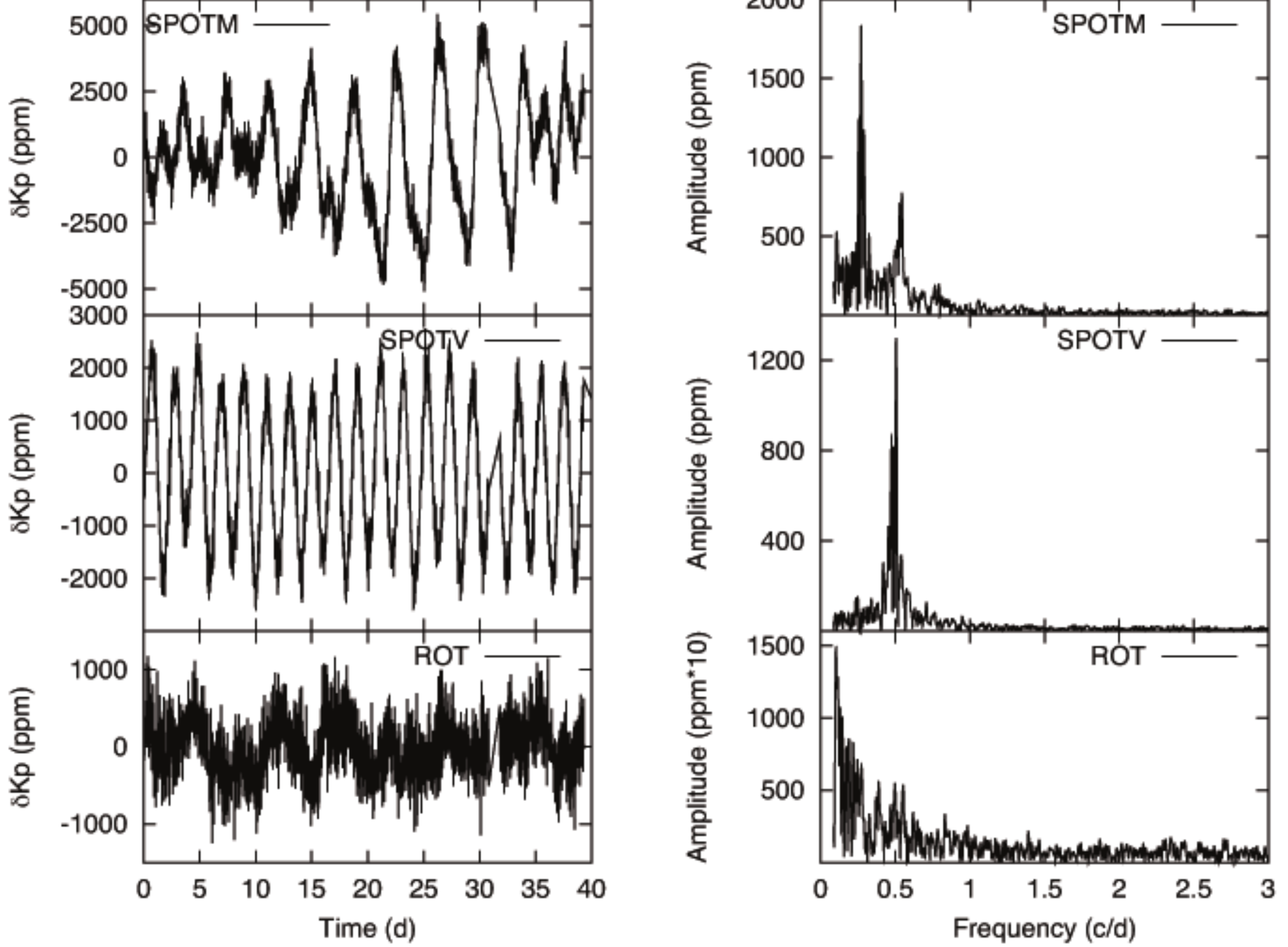}
\caption{Light curves (left panels) and Fourier Transforms (right panels) for three representative rotating, spotted stars.
 KIC3545661 (top row) is an example of a SPOTM star with multiple spots and frequencies, KIC4276984 (middle row) 
 is a SPOTV star with a single dominant period, and KIC3248536 is a ROT star with multiple frequencies below 1 c~$d^{-1}$.  
  \label{fig9}}
\end{figure}

\clearpage

\begin{figure}
\epsscale{.80}
\plotone{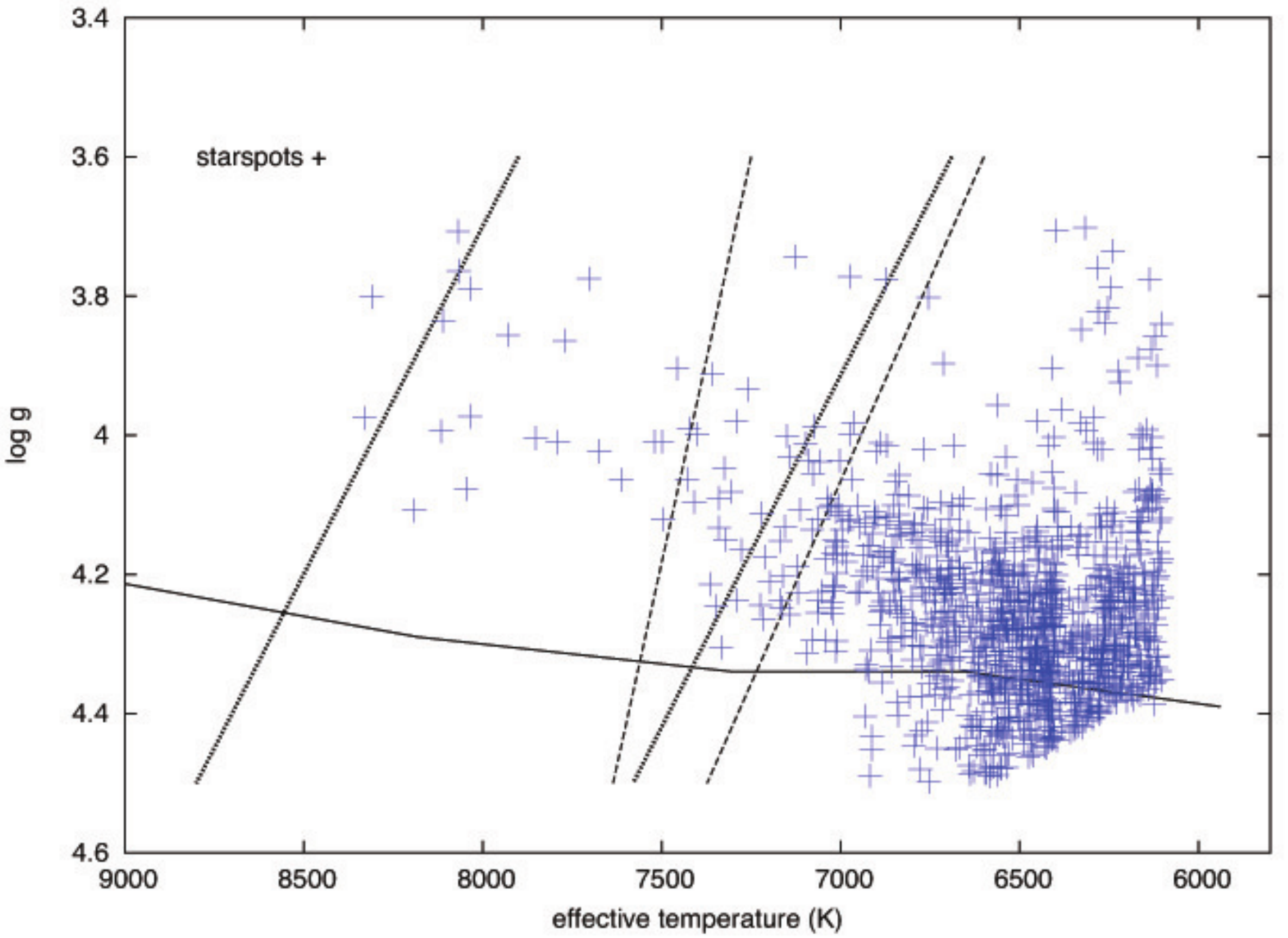}
\caption{Location of the rotating, spotted stars from our sample in the $T_{\rm eff }$, $\log$~g diagram. The ground based delta 
Scuti (thick dotted lines) and $\gamma$~Dor (thin dashed lines) instability strips are indicated, along with the zero-age 
main sequence (solid line). These stars are almost all cooler than the instability strips, as one would expect, 
since stellar activity (spots) become more prevelant with cooler $T_{\rm eff}$ values (and deeper envelope convection zones). 
The cutoffs at 6200 K and and in $\log$~g are observational selection effects of our sample.  \label{fig10}}
\end{figure}

\clearpage

  \begin{deluxetable}{lcccl}
\tabletypesize{\scriptsize}
\tablecaption{Rotating stars found in our survey}
\tablewidth{0pt}
\tablehead{
\colhead{KIC {\#}} & \colhead{$K_p$} & \colhead{${\rm T}_{\rm eff}$} & \colhead{log g} & \colhead{class} 
}
\startdata 
   1160919 &  14.2 &   6160 &  4.0 &  SPOTV\\
   1296334 &  14.9 &   6670 &  4.1 &  ROT\\
   1572948 &  14.5 &   6560 &  4.1 &  VAR\\
   1720794 &  14.5 &   6280 &  4.1 &  ROT\\
   2975747 &  14.0 &   6430 &  4.4 &  VAR\\
   2985386 &  14.0 &   6450 &  4.2 &  ROT\\
  \enddata
\tablecomments{Table 7 is published in its entirety in the 
electronic edition of the {\it Astronomical Journal}.  A portion is 
shown here for guidance regarding its form and content.
The ${\rm T}_{\rm eff}$ and {$\log$~g} values are
rounded from the {\it Kepler} input catalog.}
\end{deluxetable}

\clearpage

\subsection{Comparison of $\gamma$~Dor, $\delta$~Sct, and hybrid candidate star discovery rates} 

In this section we discuss the discovery rate of our pulsating stars, comprised of $\gamma$~Dor, $\delta$~Sct,  and 
hybrid star candidates.
Our $\gamma$~Dor candidate discovery rate ($\equiv \gamma$~Dor stars/pulsating stars) is about 64{\%}, 
the $\delta$~Sct candidate star 
discovery rate is 26{\%}, and hybrid star candidates comprise the remaining 10{\%}. 
The magnitude distribution of our $\gamma$~Dor stars has a peak at magnitude 14.3 and 
only 18 stars are fainter than magnitude 15 (see Figure 10). 
This magnitude distribution is similar to the magnitude distribution of the entire sample, however. 
The magnitude distribution of our $\delta$~Sct candidate stars is essentially flat between magnitude 14.0 
and 14.8 (see Figure 10), while only 7 are magnitude 15 or fainter (all of these are HADS stars).
There is also a dearth of faint hybrid stars in our sample.
 We find that 22 of the 32 hybrid stars lie between magnitudes 14.0 and 14.5 (see Figure 10),  
 plus one bright star at magnitude 10.3. 
 These numbers imply that we are seeing selection effect behavior, as we discuss more shortly. 

 One selection effect issue already mentioned is that almost half of our $\delta$~Sct star candidates 
 are HADS stars, so we probably are not discovering all of the low amplitude candidates. 
 This would be exacerbated if many of these stars have periods less than 2 hours; the 
 undersampling of the light curve with long-cadence data would reduce the inferred pulsation 
 amplitude. 
 Given the faintness of our stars, low amplitude modes -- especially if they are undersampled --  may well 
 have their observed amplitudes reduced to below the detection limit of {\it Kepler}. 
 We also examined the $T_{\rm eff}$ distribution of our target stars, and the bulk of our stars (see Figure 2) 
 lie between 6200 and 7000~K, which is quite a bit cooler than the Uytterhoeven et al. (2011) study, where most stars
 lie between 6500 and 8400~K and Balona and Dziembowski (2011; 6600 to 8900~K) and 
 Balona et~al. (2011; 6300 to 7200~K). 
 The cooler temperature stars in our sample would bias us towards detecting 
 $\gamma$~Dor stars rather than $\delta$~Sct or hybrid stars. 
 
 Finally, we examined the possibility that the falloff with magnitude is due to the fainter stars 
 being far enough away to lie between spiral arms in our Galaxy, where we would expect fewer stars. 
 We considered stars with luminosities between {$2 L_{\odot}$} (cool $\gamma$~Dor) and 
 $10 L_{\odot}$ (hot $\delta$~Sct). 
 The distance modulus lies between 10.91 and 12.66 magnitudes (NGC 6791 has a 
 distance modulus of 13.36 mag). 
 Brunthaler et al. (2011) present a schematic map of our Galaxy (see their Figure 2). 
 If we overlay our implied distances (1.5 to 3.4 kpc), they may lie in a space between spiral 
 arms, whereas NGC 6791 (4.7 kpc) lies in the next arm out. 
 Thus, it is possible that the relative numbers of variable stars is being affected by the 
 different sample volumes for the more luminous versus less luminous stars between spiral arms.
 The $\gamma$~Dor stars, being less luminous, would be more likely to lie within our spiral 
 arm, where there are more stars. 
 On the other hand the more luminous $\delta$~Sct stars in our sample may be far enough away to lie in 
 the interarm space where there are fewer stars. 
 This combination of effects appears to make it less likely to discover faint, low amplitude 
 $\delta$~Sct and hybrid stars, and would explain the relative lack of these stars in our sample.
 
 \begin{figure}
\epsscale{.80}
\plotone{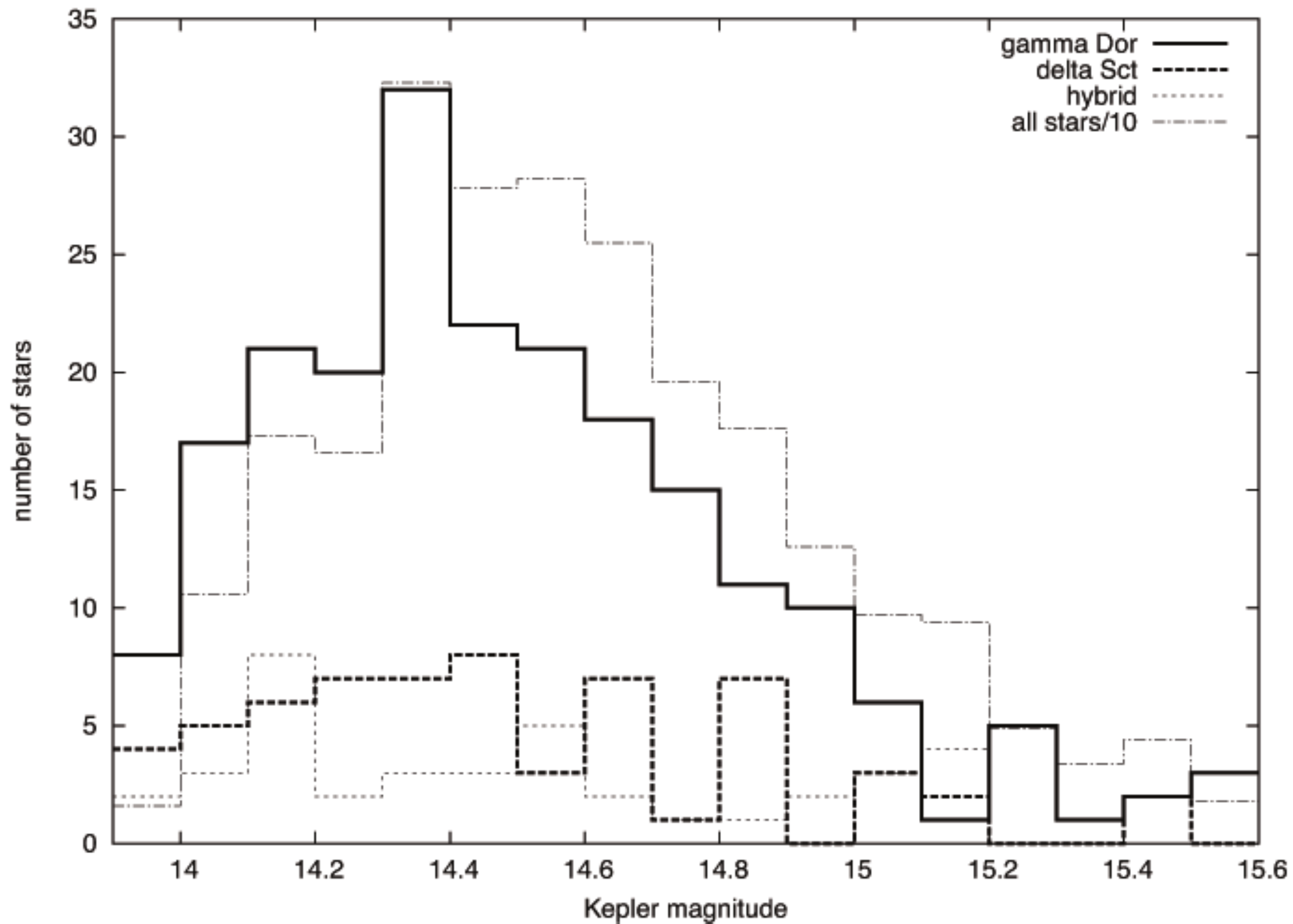}
\caption{Histogram of the $\gamma$~Dor, $\delta$~Sct, and hybrid candidate stars binned by {\it Kepler} magnitude ($K_p$). 
The $\gamma$~Dor star candidates (solid line) have a distribution that resembles the overall sample (dotted line). 
The $\delta$~Sct star candidates (long-dashed line), and hybrid star candidates (short dashed line) do not have obvious peaks 
in their distributions, but both have very few stars fainter than $K_p$ magnitude 15.   \label{fig11}}
\end{figure}

\clearpage

\section{Summary} 

In this study, we examined the light curves of 2768 stars, mostly between magnitudes 14 and 
15, that were selected with temperatures between 6200 and 8200 K, placing them in or near 
the known $\gamma$~Dor and $\delta$~Sct instability strips. 
We found 1531 stars that exhibited some sort of variability in their light curves, of which 207 
are $\gamma$~Dor candidate stars, 84 are $\delta$~Sct candidate stars, and 32 are hybrid
 candidate stars. 
 The temperature distribution of our $\gamma$~Dor, $\delta$~Sct, and hybrid stars is similar to that of 
 Balona et al. (2011), Balona {\&} Dziembowski (2011) and Uytterhoeven et al. (2011). 
 Almost all of our $\gamma$~Dor candidates lie between 6100 and 7500 K, which is similar to 
 the values of Balona et al. (2011), but cooler than that of Uytterhoeven et al. (2011). 
 Our sample has more stars below the red edge of the ground based instability strip
($\sim 50${\%}) than the other two studies, which is the result of the effective temperature distribution of our sample.
 Our hybrid star candidates are scattered nearly uniformly from 6100 to 8000 K, and this is 
 consistent with Uytterhoeven et al. (2011), except that their temperature range is 6600 to 8200 K. 
 Finally, our $\delta$~Sct candidate sample has far more cool stars (below about 6700 K) than Balona {\&}
 Dziembowski (2011) or Uytterhoeven et al. (2011), but if we remove the our HADS candidates (recall that
 these may well be rotating, spotted stars), then many of our $\delta$~Sct candidates lie in or near the
 ground based $\delta$~Sct instability strip. 
 Guzik~et~al. (2013, 2014) found a few constant (no FT peaks greater than 20 ppm) stars within the
 ground-based instability strips as well.

 We also found 76 binary systems and 1132 stars with low frequency variations that we attribute 
 to rotation or some other phenomenon. We note that nine of the rotating stars have $T_{\rm eff}$ 
 values over 8000K and have properties similar to the ones discussed by Balona (2013). 
 We compared the relative detection rates of $\gamma$~Dor, $\delta$~Sct, and hybrid star candidates 
 and find that our sample is dominated by $\gamma$~Dor candidates, at 64{\%}. 
 We also note that many of our $\delta$~Sct star candidates are HADS stars, which are easier to 
 detect due to their large amplitudes and longer periods than the average $\delta$~Sct star. 
 Short cadence data might help detect more $\delta$~Sct stars. 
 We found 323 $\gamma$~Dor, $\delta$~Sct, and hybrid stars, but in all three cases 
 the number of discoveries falls off rapidly at a magnitude fainter than 15.0, implying that selection effects 
 are limiting the number of faint stars we can discover. 
 One possibility is that the fainter stars are far enough away to lie between spiral arms in our Galaxy, 
 where there would be fewer stars.

\acknowledgments

We acknowledge support from the {\it Kepler} Guest Observer program. 
Part of this work was funded by NASA grants Cycles 1-4. 
This paper includes data collected by the {\it Kepler} mission. Funding for the {\it Kepler} mission is provided by
the NASA Science Mission Directorate.

K.U. acknowledges support from the Spanish National Plan of R{\&}D for 2010, project 
AYA2010-17803. 
The research leading to these results has received funding from the European Community's Seventh 
Framework Programme (FP7/2007-2013) under grant agreement no. 269194.
This project benefited from Project FP-7-PEOPLE-IRSES:ASK Np. 269194. 
We acknowledge fruitful discussions with Andrezj Pigulski, Joanna Molenda-Zakowicz, 
Patrick Gaulme, and Gerald Handler. 
We thank the anonymous referee for their careful reading of the manuscript and comments, 
which greatly improved this paper.



{\it Facilities:} \facility{Kepler}.

\clearpage

\clearpage

\begin{deluxetable}{lccclccc}
\tabletypesize{\scriptsize}
\tablecaption{$\gamma$~Dor candidate stars found in our survey}
\tablewidth{0pt}
\tablehead{
\colhead{KIC {\#}} & \colhead{$K_p$} & \colhead{${\rm T}_{\rm eff}$} & \colhead{log g} & \colhead{class} &
\colhead{Freq. Range}&\colhead{Ampl. high}&\colhead{Freq. high}}
 \startdata
                   &          &           &             &            & (${\rm d}^{-1}$) & (ppm) & (${\rm d}^{-1}$) \\
   2167444 & 14.1  & 7140 & 4.1 & MULT &  $0.5 - 5.2$       &  1730  &    0.8082            \\
   2448307 & 14.0  & 7350 & 3.9 & MULT &  $0.2 - 3.5$        &  1390  &   1.2516            \\
   2579595 & 14.1  & 7150 & 4.1 & ASYM &  $0.7 - 1.7$       &   5175  &    1.2205           \\
   2581964 & 14.0  & 7410 & 4.2 & ASYM &  $0.4 - 1.5$       &   5914  &   0.5389            \\
   2857178 & 14.6  & 7440 & 4.2 & ASYM &  $0.9 - 2.0$       &     392  &   1.7313            \\
   2974858 &  14.4 &   7010 &  4.2 &  MULT& $0.75 - 3.1$   &       49 &    2.0287            \\
   2975214 &  14.0 &   6860 &  4.2 &  MULT& $0.25 - 0.5$  &       69  &   0.3131            \\
   2985526 &  14.4 &   6180 &  4.3 &  MULT& $.25 - 1.2$    &        53  &   0.3737           \\
   2996232 &  14.1 &   7180 &  4.2 &  ASYM& $0.35 - 1.5$  &      420 &   0.9369           \\
   3248951 &  14.3 &   7270 &  4.3 &  SYM   & $0.25 - 3.0$  &       400 &   1.4643           \\
   3556003 &  14.3 &   6710 &  4.1 &  SYM   & $0.75 - 1.25$&    6100 &    0.9381         \\
   3660010 &  14.2 &   6360 &  4.4 &  SYM   & $0.35 - 1.0$  &    3500  &   0.3902         \\
   3853742 &  14.1 &   7280 &  4.2 &  SYM   & $0.25 - 0.75$&      335  &   0.4416         \\
   3947031 &  14.1 &   7270 &  4.2 &  SYM   & $0.75 - 1.5$   &     432  &   0.9082         \\
   3964570 &  14.0 &   6760 &  4.3 &  SYM   & $0.75 - 1.25$ &   1140  &   1.0841         \\
   3966800 &  14.7 &   6680 &  3.9 &  MULT & $0.25 - 1.5$   &      101  &   0.3294        \\
   3967081 &  14.2 &   6700 &  4.2 &  SYM    & $0.25 - 2.75$&    3330  & 1.2850         \\
   4155654 &  15.4 &   7360 &  3.9 &  MULT  & $0.20 - 0.75$&     132   & 0.2208        \\
   4281025 &  14.0 &   6300 &  4.4 &  MULT  & $0.20 - 1.0$   &     390  &  0.4136       \\
   4346807 &  14.9 &   6480 &  4.3 &  SYM    & $0.35 - 0.75$ &     300  &  0.4596       \\
   4365676 &  14.5 &   6450 &  4.3 &  SYM    & $0.5 - 1.4$      &    7350 & 0.6464        \\
   4372107 &  15.0 &   7660 &  4.2 &  SYM    & $1.3 - 1.8$      &    3500 & 1.6250        \\
   4554290 &  14.3 &   6620 &  4.4 &  MULT  & $0.25 - 1.2$   &     1530 & 0.3558       \\
   4561941 &  14.6 &   6550 &  3.8 &  MULT  & $0.25 - 1.0$   &      1390 & 0.6437      \\
   4566563 &  14.2 &   7090 &  4.2 &  MULT  & $0.3 - 1.5$     &         250 & 0.3435      \\
   4568058 &  14.4 &   7380 &  4.2 &  SYM     & $0.8 - 1.3$    &       8000 & 1.0794      \\
   4568087 &  14.8 &   7350 &  4.1 &  SYM     & $0.7 - 1.0$    &       2200 & 0.8505      \\
   4572083 &  14.9 &   6880 &  4.3 &  SYM     & $0.8 - 1.2$    &     13000 & 0.8925      \\
   4830647 &  14.0 &   7270 &  4.1 &  MULT   & $0.9 - 4.0$    &       1600 & 1.5175      \\
   4840978 &  14.6 &   6410 &  4.3 &  MULT   & $0.5 - 1.3$    &         150  & 0.6250     \\
   4841423 &  14.6 &   6820 &  4.1 &  MULT   & $0.6 - 1.0$    &            75  & 0.6529     \\
   4843904 &  14.2 &   6410 &  4.3 &  MULT   & $0.2 - 0.8$    &            90  & 0.2967    \\
   4930756 &  15.2 &   7280 &  4.3 &  MULT   & $0.25 - 1.5$  &        1100 & 0.3350    \\
   4934217 &  14.2 &   7560 &  3.8 &  ASYM   & $0.4 - 1.0$    &        7960 & 0.8809    \\
   5097566 &  14.3 &   6640 &  4.3 &  MULT   & $0.5 - 1.6$     &          150 & 0.5071    \\   
   5108395 &  14.2 &   6440 &  4.4 &  SYM     & $0.3 - 1.2$      &          900 & 0.3929   \\   
   5113089 &  14.7 &   6520 &  4.4 &  MULT   & $0.7 - 2.0$     &             70 & 0.8333   \\
   5120569 &  14.1 &   6500 &  4.0 &  MULT   & $0.2 - 1.0$     &           165 & 0.2310  \\     
   5123134 &  14.3 &   6510 &  4.4 &  MULT   & $0.2 - 3.0$     &         1250 & 0.4151 \\
   5253199 &  15.3 &   7300 &  4.1 &  ASYM  & $0.6 - 1.5$     &       13,150 & 0.6882 \\
   5254176 &  14.7 &   6880 &  4.1 &  ASYM  & $0.4 - 1.6$     &       15,780 & 0.7878 \\
   5280974 &  14.4 &   7010 &  4.2 &  SYM     & $0.75 - 1.75$ &      14,000 & 1.5210 \\
   5281931 &  15.0 &   6860 &  4.2 &  MULT   & $0.25 - 1.25$ &            240 & 0.3551 \\
   5342935 &  14.0 &   6760 &  4.4 &  MULT   & $0.6 - 2.0$      &          1905 & 0.6849 \\
   5357521 &  14.8 &   7490 &  4.1 &  MULT   & $0.5 - 4.0$      &          2100 & 2.0832 \\
   5471623 &  14.1 &   6770 &  4.1 &  MULT   & $0.2 - 2.2$      &          2300 & 0.2488 \\
   5524370 &  14.5 &   7710 &  4.0 &  SYM     & $0.4 - 0.8$      &          6570 & 0.6143 \\
   5529141 &  14.3 &   7270 &  4.2 &  SYM     & $1.1 - 2.5$     &           2935 & 1.6944 \\
   5531052 &  14.2 &   6850 &  4.2 &  MULT   & $0.5 - 1.0$      &          1950 & 0.7400 \\
   5541405 &  14.6 &   6400 &  4.4 &  MULT   & $0.25 - 0.75$ &             220 & 0.3271 \\
   5543818 &  14.5 &   6290 &  4.3 &  SYM     & $0.2 - 0.4$      &           1250 & 0.2033 \\
   5561007 &  14.2 &   7690 &  4.2 &  MULT   & $1.0 - 6.0$     &            6915 & 1.2411 \\
   5615977 &  14.9 &   6820 &  4.2 &  SYM     & $0.25 - 0.7$    &           2440 & 0.3213 \\
   5633114 &  14.8 &   7090 &  4.2 &  SYM     & $0.5 - 2.75$   &              750 & 1.6098 \\
   5696153 &  14.1 &   7270 &  4.0 &  MULT  & $0.25 - 1.5$    &            2600 & 0.3022 \\
   5717449 &  14.9 &   7480 &  4.1 &  MULT  & $0.25 - 4.25$  &            1200 & 1.8505 \\
   5723583 &  14.7 &   6460 &  4.2 &  MULT  & $0.3 - 0.7$      &             2010 & 0.5710 \\
   5724633 &  14.2 &   6410 &  3.9 &  MULT  & $0.8 - 4.0$      &             2885 & 2.6317 \\
   5869220 &  14.4 &   6680 &  4.2 &  MULT  & $0.3 - 0.9$      &             1815 & 0.4586 \\
   5878159 &  14.6 &   7330 &  4.0 &  ASYM  & $0.3 - 1.0$     &           13,905 & 0.3690 \\
   5879583 &  14.3 &   7400 &  4.0 &  ASYM  & $2.3 - 2.5$     &           67,390 & 2.3872 \\
   5880445 &  14.3 &   7390 &  4.3 &  MULT  & $1.1 - 2.6$     &              6280 & 1.3180 \\
   5888202 &  14.1 &   6360 &  4.2 &  MULT  & $0.2 - 0.7$     &                400 & 0.2103 \\
   5893541 &  14.5 &   6560 &  4.3 &  MULT  & $0.5 - 1.25$  &              2025 & 0.5944 \\
   5899833 &  14.5 &   6520 &  4.0 &  ASYM  & $0.4 - 1.25$  &             9800 & 0.4798 \\
   5954775 &  14.6 &   6540 &  4.2 &  MULT  & $0.2 - 0.6$     &             3445 & 0.2962 \\
   5954946 &  14.3 &   6680 &  4.3 &  MULT  & $0.25 - 1.0$   &                 90 & 0.4369 \\
   5957185 &  14.3 &   7080 &  4.1 &  MULT  & $0.3 - 1.8$    &              1845 & 0.6106 \\
   5960912 &  14.7 &   6840 &  4.4 &  MULT  & $0.5 - 1.0$    &              1525 & 0.6642\\
   6035618 &  15.7 &   7400 &  4.1 &  ASYM  & $0.75 - 1.25$ &        249,840& 1.0936\\
   6048208 &  14.9 &   6880 &  4.3 &  MULT  & $0.25 - 1.0$   &             1230 & 0.3151 \\
   6119163 &  14.6 &   7350 &  4.0 &  MULT  & $1.0 - 3.5$     &              1425 & 1.1344 \\
   6128330 &  14.0 &   7450 &  4.0 &  ASYM  & $0.25 - 1.1$   &           21,925 & 0.4553 \\
   6130543 &  14.7 &   6840 &  4.3 &  MULT  & $0.25 - 1.0$   &              1690 & 0.4293 \\
   6131093 &    9.3 &   6640 &  4.4 &  MULT  & $1.25 - 4.0$   &           17,090 & 1.5923 \\
   6145991 &  14.1 &   6210 &  4.2 &  MULT  & $1.0 - 2.0$     &               1615 & 1.1428 \\
   6187850 &  14.2 &   6260 &  4.2 &  SYM    & $0.25 - 0.75$ &                930 & 0.3526 \\
   6197138 &  14.5 &   6600 &  4.3 &  SYM    & $0.25 - 0.75$ &              2750 & 0.3575 \\
   6210324 &  14.5 &   7580 &  4.0 &  MULT  & $0.20 - 5.0$   &                625  & 2.4603 \\
   6220939 &  14.4 &   6580 &  4.5 &  MULT  & $0.20 - 1.5$  &                 105 & 0.2336 \\
   6230552 &  14.6 &   6600 &  4.3 &  ASYM   & $0.75 - 2.0$   &            17,000 & 1.5467\\
   6279848 &   8.9  &   N.A.   &   N.A.   &  MULT   & $0.75 - 1.6$  &           109,900 & 1.1552\\
   6301965 &  14.3 &   7080 &  4.2 &  MULT   & $0.5 - 3.7$     &                 225 & 0.9065\\
   6302177 &  14.3 &   6840 &  4.2 &  SYM    & $1.25 - 3.25$ &                 175 & 1.4883 \\
   6302332 &  14.0 &   6270 &  4.1 &  SYM     & $0.25 - 0.80$  &               2300 & 0.3435\\
   6302576 &  14.4 &   7200 &  4.1 &  MULT   & $0.25 - 3.1$   &                  370 & 1.0724\\
   6302810 &  14.3 &   6900 &  4.1 &  MULT   & $0.25 - 1.0$   &                    65 & 0.2804\\
   6305211 &  14.4 &   7040 &  4.2 &  ASYM  & $0.25 - 0.75$ &                 430 & 0.4533\\
   6311979 &  14.5 &   6150 &  4.3 &  ASYM  & $0.20 - 0.60$ &               2300 & 0.2617 \\
   6370357 &  14.5 &   6180 &  4.3 &  SYM   & $0.25 - 0.60$ &                3400 & 0.2991\\
   6373589 &  14.2 &   7250 &  3.9 &  MULT & $0.25 - 1.25$  &                   53 & 0.3271 \\
   6373976 &  14.5 &   6900 &  4.3 &  MULT & $0.25 - 1.0$    &                   65 & 0.2804 \\
   6388558 &  14.3 &   6820 &  4.1 &  MULT & $0.9 - 4.6$      &                 800 & 3.1682 \\
   6497099 &  14.8 &   6880 &  4.2 &  ASYM & $0.25 - 0.9$    &               6650 & 0.4136 \\
   6509590 &  14.7 &   7690 &  4.1 &  SYM    & $1.8 - 4.6$     &                  725 & 2.4369 \\
   6509962 &  14.6 &   6700 &  4.2 &  SYM    & $0.4 - 1.6$     &                  425 & 0.4883 \\
   6515882 &  14.5 &   7060 &  4.3 &  SYM    & $0.4 - 1.5$     &              11875 & 1.3214 \\
   6521724 &  14.4 &   6340 &  4.3 &  SYM    & $0.25 - 0.75$ &               1610  & 0.3481 \\
   6526666 &  14.2 &   7030 &  4.2 &  SYM    & $0.3 - 0.8$    &                   160 & 0.5000 \\
   6547130 &  14.1 &   6460 &  4.0 &  ASYM  & $0.8 - 2.5$    &               15000 & 1.1869  \\
   6580506 &  14.5 &   7070 &  4.0 &  MULT  & $0.25 - 2.5$  &                  1510 & 0.5838 \\
   6599573 &  14.4 &   7290 &  4.2 &  MULT  & $1.0 - 4.3$    &                       60 & 2.1429 \\
   6628926 &  14.1 &   6220 &  4.1 &  SYM    & $1.3 - 2.5$    &                   3200 & 2.0818 \\
   6691019 &  15.0 &   6950 &  4.3 &  ASYM  & $1.1 - 3.6$   &                   2000 & 1.2600 \\
   6700412 &  14.8 &   7430 &  4.0 &  SYM    & $0.5 - 0.8$    &                  1475 & 0.6117 \\
   6708546 &  14.3 &   6580 &  4.4 &  MULT  & $0.3 - 1.5$   &                        75 & 0.5093 \\
   6721297 &  14.8 &   6550 &  4.3 &  ASYM & $0.75 - 1.8$  &                  5000 & 1.7500 \\
   6764033 &  14.0 &   6920 &  4.2 &  SYM   & $0.2 - 0.4$    &                     135  & 0.2617 \\
   6777294 &  14.5 &   6640 &  4.3 &  SYM   & $0.25 - 0.6$   &                   2250 & 0.2550 \\
   6778212 &  14.4 &   7360 &  4.1 &  MULT & $0.3 - 1.1$   &                        35 & 0.3598  \\
   6778479 &  14.9 &   7090 &  4.2 &  MULT  & $0.3 - 1.1$  &                        55 & 1.4405 \\
   6791943 &  15.0 &   6310 &  4.4 &  SYM    & $0.2 - 0.8$  &                     2200 & 0.2500 \\
   6803941 &  14.2 &   6100 &  3.7 &  MULT  & $0.2 - 0.75$ &                      110 & 0.2804 \\
   6870063 &  14.7 &   6720 &  4.3 &  MULT  & $0.3 - 1.25$ &                        90 & 0.3621 \\
   6871727 &  14.4 &   7010 &  4.2 &  MULT  & $0.2 - 0.7$   &                       135 & 0.3435 \\
   6878998 &  14.3 &   6200 &  4.3 &  MULT  & $0.35 - 2.0$ &                        80 & 0.4136 \\
   6937000 &  14.7 &   6120 &  4.3 &  SYM    & $0.5 - 1.25$ &                    2000 & 0.6117 \\
   6954657 &  14.7 &   6860 &  4.3 &  MULT  & $ 0.25 - 4.0$ &                       32 & 2.0000 \\
   6964677 &  14.2 &   6760 &  4.2 &  MULT  & $1.5 - 3.5$   &                        34 & 2.8262 \\
   6965624 &  15.2 &   6560 &  4.4 &  MULT  & $0.9 - 3.4$   &                      650 & 1.9369 \\
   7019480 &  14.6 &   6220 &  4.3 &  SYM    & $0.3 - 1.1$   &                     7200 & 0.3388 \\
   7031347 &  14.6 &   6910 &  4.3 &  SYM    & $0.3 - 0.5$   &                       160 & 0.4136 \\
   7098418 &  15.5 &   7350 &  4.2 &  SYM    & $1.5 - 3.5$   &                     2720 & 1.7500 \\
   7187175 &  14.9 &   7070 &  4.1 &  MULT  & $0.2 - 1.5$   &                   38730 & 0.2874 \\
   7191437 &  14.3 &   6900 &  4.5 &  SYM    & $0.5 - 2.5$   &                          90 & 0.6285 \\
   7191683 &  14.5 &   7660 &  4.1 &  SYM    & $1.0 - 1.5$   &                        750 & 1.3902 \\
   7220299 &  14.6 &   6300 &  4.2 &  SYM    & $0.25 - 0.6$ &                        410 & 0.2804 \\
   7222218 &  14.6 &   6160 &  4.0 &  SYM    & $ 1.0 -2.3$   &                        175 & 1.1402 \\
   7296779 &  14.6 &   6420 &  4.3 &  MULT  & $0.2 - 2.5$  &                         190 & 0.5187 \\
   7297626 &  14.7 &   6320 &  4.3 &  MULT  & $0.25-1.25$ &                        85  & 0.2967 \\
   7298757 &  14.3 &   6110 &  4.2 &  MULT  & $0.25 -1.5$ &                         35 & 0.2804 \\
   7445690 &  14.8 &   6950 &  4.2 &  MULT  & $0.4 - 1.4$  &                         80 & 0.4717 \\
   7448050 &  11.8 &   7390 &  4.1 &  ASYM  & $1.0 - 1.7$ &                611,990 & 1.1430 \\
   7456325 &  14.3 &   6540 &  4.5 &  SYM    & $0.25 - 1.0$ &                      800 & 0.2780 \\
   7512583 &  14.4 &   6380 &  4.2 &  SYM    & $1.7-3.5$   &                      3600 & 1.7383 \\
   7523217 &  14.3 &   7080 &  4.3 &  MULT  & $0.75-2.1$ &                       100 & 0.8225 \\
   7534444 &  14.3 &   6290 &  4.4 &  MULT  & $0.25-0.75$ &                       80 & 0.4486 \\
   7538543 &  14.9 &   6870 &  4.2 &  SYM    & $0.2-4.8$ &                          900 & 0.3201 \\
   7552602 &  15.0 &   8080 &  3.9 &  SYM    & $3.25-3.6$ &                       370 & 3.5476 \\
   7553070 &  14.7 &   6410 &  4.1 &  SYM    & $0.45-1.7$ &                     1510 & 1.0030 \\
   7581111 &  15.2 &   6540 &  4.2 &  SYM    & $0.5-1.7$   &                     2590  & 0.5234 \\
   7685157 &  14.0 &   7270 &  4.3 &  SYM    & $0.75-1.5$ &                    8940 & 0.8562 \\
   7685306 &  14.3 &   7900 &  4.0 &  MULT  & $0.75-4.0$ &                 12,865 & 0.7874 \\
   7687374 &  14.5 &   8020 &  3.9 &  ASYM  & $0.4-1.0$  &                     5540 & 0.8435 \\
   7693255 &  14.0 &   6840 &  4.2 &  MULT  & $0.4-1.5$  &                        150 & 0.3971 \\
   7756660 &  14.1 &   7030 &  4.2 &  MULT  & $0.25-1.2$ &                       140 & 0.7500 \\
   7765798 &  14.8 &   7200 &  3.9 &  MULT  & $0.3-1.5$  &                           68 & 0.6869 \\
   7767565 &   9.3 &    N.A.   &   N.A.   &  MULT  & $1.1-4.2$  &                   30,065 &1.4125 \\
   7771914 &  14.1 &   6800 &  4.2 &  SYM    & $0.65-2.7$   &                      125 & 0.6833 \\
   7821231 &  14.0 &   6690 &  4.2 &  MULT  & $0.25-1.1$ &                       500 & 0.3271 \\
   7834518 &  14.5 &   7080 &  4.2 &  MULT  & $0.75-4.25$ &                     125 & 2.0818 \\
   7891007 &  14.1 &   6420 &  4.2 &  MULT  & $0.25-1.6$  &                      145 & 0.3037 \\
   7903015 &  15.0 &   6680 &  4.3 &  MULT  & $0.25-1.75$ &                     110 & 0.4265\\
   7907511 &  14.7 &   6170 &  4.2 &  SYM     & $0.75-1.75$ &                  4100 & 0.9720 \\
   7908851 &  14.8 &   7180 &  4.1 &  SYM     & $1.8-3.25$   &                  2200 & 2.6869 \\
   7968267 &  14.3 &   7260 &  4.2 &  SYM     & $0.3-1.1$     &                  4670 & 0.4860 \\
   7988596 &  14.1 &   6490 &  4.2 &  SYM     & $0.4-1.2$     &                    570 & 0.5117 \\
   8028916 &  14.1 &   7060 &  4.1 &  ASYM  & $0.35-1.4$   &              26,095 & 0.6552 \\
   8035262 &  14.3 &   7290 &  4.1 &  SYM     & $0.75-3.0$  &                 2075  & 1.3196 \\
   8113557 &  14.5 &   6720 &  4.1 &  ASYM  & $1.2-1.7$    &                  3900  & 1.3750 \\
   8124401 &  15.2 &   6310 &  3.9 &  ASYM  & $1.0-1.25$  &              11,000  & 1.1075 \\
   8189115 &  14.3 &   6270 &  4.2 &  MULT  & $0.7-2.0$    &                    100  & 0.7734 \\
   8189504 &  14.5 &   6580 &  4.4 &  SYM    & $0.75-2.1$  &                    700  & 1.0250 \\
   8221845 &  14.4 &   7470 &  4.1 &  SYM    & $1.9-3.3$    &                  1250  & 2.4439 \\
   8221986 &  14.6 &   7180 &  4.1 &  MULT  & $0.9-2.7$    &                  1345 & 1.2349 \\
   8222685 &   8.9 &    N.A.   &   N.A.   &  ASYM  & $1.1-2.1$    &             120,765 & 1.2218 \\
   8309369 &  14.9 &   7260 &  4.0 &  SYM     & $3.75-4.25$&                    250 & 3.9952 \\
   8313084 &  14.8 &   6500 &  4.4 &  MULT   & $0.25-1.0$  &                    110 & 0.2804 \\
   8323022 &  14.4 &   6280 &  4.2 &  MULT   & $0.2-2.0$    &                     120 & 0.4416 \\
   8327168 &  14.5 &   6300 &  4.3 &  SYM     & $0.45-1.2$  &                    1600 & 0.5050 \\
   8359028 &  14.4 &   6740 &  4.3 &  SYM     & $0.35-1.0$  &                    1100 & 0.7617 \\
   8398162 &  14.3 &   6530 &  4.3 &  SYM     & $0.5-1.5$    &                      200 & 0.8037 \\
   8410612 &  14.2 &   6640 &  4.2 &  SYM     & $0.3-1.5$    &                      600 & 0.3785 \\
   8415383 &  14.0 &   7360 &  4.1 &  SYM     & $1.0-4.5$    &                      410 & 3.0327 \\
   8478472 &  14.4 &   7390 &  4.1 &  ASYM  & $1.1-2.3$    &                    8600 & 2.1098 \\
   8481328 &  14.7 &   7120 &  4.1 &  SYM     & $0.5-1.2$    &                      160 & 0.6495 \\
   8508852 &  14.5 &   6760 &  4.2 &  SYM     & $1.0-5.0$    &                      210 & 2.6551 \\
   8526232 &  14.1 &   6770 &  4.2 &  SYM     & $1.0-3.3$    &                   2100 & 1.9556 \\
   8638619 &  14.2 &   7340 &  4.3 &  SYM     & $1.25-2.75$ &                 2520 & 2.2220 \\
   8767298 &  14.6 &   6640 &  4.5 &  SYM     & $0.3-1.9$    &                     220 & 0.7874 \\
   8804158 &  14.1 &   7020 &  3.9 &  SYM     & $0.75-2.0$ &                     175 & 0.7967 \\
   8814047 &  14.3 &   6860 &  4.3 &  SYM     & $0.3-1.3$   &                      130 & 0.5888 \\
   9012615 &  15.2 &   7240 &  4.3 &  MULT  & $1.3-3.0$    &                        46 & 1.4439 \\
   9050484 &  14.1 &   6420 &  4.0 &  SYM    & $0.5-3.25$  &                      130 & 0.5374 \\
   9053846 &  14.0 &   6300 &  4.1 &  ASYM & $0.9-1.5$     &                    7300 & 1.3037 \\
   9073315 &  14.6 &   7320 &  4.3 &  SYM    & $0.7-1.6$    &                49,755 & 0.76601\\
   9113086 &  14.8 &   6540 &  4.1 &  SYM    & $0.45-1.25$&                   1500 & 0.5864 \\
   9149977 &  14.4 &   6900 &  4.5 &  SYM    & $0.4-1.4$    &                      120 & 0.5397 \\
   9241468 &  14.5 &   6190 &  3.9 &  SYM    & $0.35-1.75$&                      110 & 0.5467 \\
   9290681 &  14.3 &   7360 &  4.1 &  SYM    & $0.6-0.8$    &                     1500 & 0.7850 \\
   9304984 &  14.3 &   6220 &  4.4 &  SYM    & $0.75-21$  &                      4400 & 0.8598 \\
   9405431 &  14.3 &   7000 &  4.2 &  SYM    & $0.25-2.1$ &                        120 & 0.3949\\
   9517548 &  15.4 &   6990 &  4.1 &  SYM    & $0.5-1.9$   &                      2235 & 0.5759 \\
   9569705 &  15.8 &   7290 &  4.0 &  SYM    & $0.35-1.2$ &                  81,730 & 0.5328 \\
   9662168 &  14.2 &   7170 &  4.1 &  ASYM  & $0.5-1.3$  &                     3000 & 0.6355 \\
   9715349 &  14.2 &   7320 &  4.1 &  ASYM  & $0.8-1.8$  &                     3960 & 1.4571 \\
   9905540 &  14.2 &   7210 &  4.1 &  SYM    & $0.5-2.25$ &                    1845 & 1.0864 \\
   9910156 &  14.3 &   7330 &  4.2 &  SYM    & $1.5-5.0$   &                       400 & 3.9836\\
   9962653 &  10.1 &   7240 &  4.3 &  ASYM & $1.4-2.5$   &                     6000 &2.2827 \\
  10015325 &  14.2 &   6860 &  4.3 &  SYM  & $0.3-1.4$   &                       220 & 0.8318 \\
  10096798 &  14.0 &   6560 &  4.3 &  SYM+ROT& $1.1-1.75$&                 230 & 1.4743 \\ 
  10683303 &    6.0 &    N.A.   &   N.A.  &  SYM+ROT&$1.25-2.5$&                  880 & 2.2617 \\
  10845049 &  14.4 &   7450 &  4.1 &  SYM  & $1.0-3.0$   &                       3345 & 1.4481 \\
  11233133 &  14.7 &   7800 &  4.1 &  ASYM& $1.5-3.5$  &                    20,000 & 1.6869 \\
  11657064 &  14.1 &   7480 &  4.0 &  ASYM& $0.4-0.8$  &                    19,200 & 0.4656 \\
  11803734 &  14.4 &   6460 &  4.2 &  SYM  & $0.35-1.2$ &                  198,360 & 0.3725 \\
  11962187 &  14.0 &   7160 &  4.0 &  ASYM& $0.3-2.2$  &                       9970 & 1.0178 \\
  \enddata
\tablecomments{``Ampl. high'' and ``Freq. high'' refer to the amplitude and
frequency of the highest amplitude mode in the FT. The ${\rm T}_{\rm eff}$ and {$\log$~g} values are
rounded from the {\it Kepler} input catalog.}
\end{deluxetable}

\clearpage


\end{document}